\documentclass[sigconf,authorversion,nonacm]{acmart}

\AtBeginDocument{%
  }

\begin{document}

\title{PromptInfuser: How Tightly Coupling AI and UI Design Impacts Designers' Workflows}



\author{Savvas Petridis}
\affiliation{%
  \institution{Google Research}
  \city{New York}
  \state{New York}
  \country{USA}}
\email{petridis@google.com}

\author{Michael Terry}
\affiliation{%
  \institution{Google Research}
  \city{Cambridge}
  \state{Massachusetts}
  \country{USA}}
\email{michaelterry@google.com}

\author{Carrie J. Cai}
\affiliation{%
  \institution{Google Research}
  \city{Mountain View}
  \state{California}
  \country{USA}}
\email{cjcai@google.com}

\renewcommand{\shortauthors}{Savvas Petridis, Michael Terry, Carrie J. Cai}



\begin{abstract}
Prototyping AI applications is notoriously difficult.
While large language model (LLM) prompting has dramatically lowered the barriers to AI prototyping, designers are still prototyping AI functionality and UI separately.
We investigate how coupling prompt and UI design affects designers' workflows. Grounding this research, we developed PromptInfuser, a Figma plugin that enables users to create semi-functional mockups, by connecting UI elements to the inputs and outputs of prompts. 
In a study with 14 designers, we compare PromptInfuser to designers’ current AI-prototyping workflow. 
PromptInfuser was perceived to be significantly more useful for communicating product ideas, more capable of producing prototypes that realistically represent the envisioned artifact, more efficient for prototyping, and more helpful for anticipating UI issues and technical constraints. 
PromptInfuser encouraged iteration over prompt and UI together, which helped designers identify UI and prompt incompatibilities and reflect upon their total solution. Together, these findings inform future systems for prototyping AI applications.
\end{abstract}

\begin{teaserfigure}
\centering
  \includegraphics[width=0.9\linewidth]{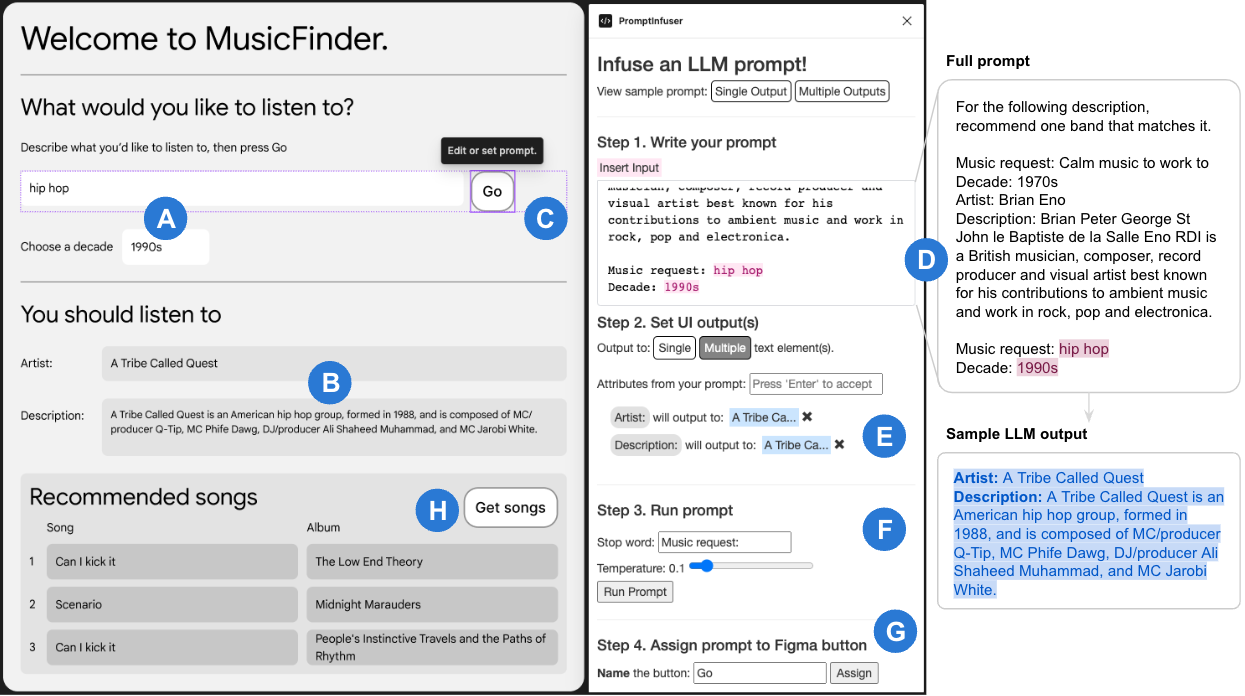}
  \caption{\textbf{PromptInfuser Plugin.} On the left is a Figma mockup for a music search application that was made semi-functional with PromptInfuser on the right. When the ``Go'' button (C) is clicked by a user, a music search prompt is called (D). The text elements in (A), in this case: ``hip hop'' and ``1990s'',  are inputted into the prompt, which is then run. Afterwards, the completion is split and mapped to the text elements in (B). To split a completion into multiple outputs, the user needs to define a one-shot example in their prompt (D) and then identify the tags to split the completion with in (E). Finally, they can assign the prompt to the ``Go'' button in (G). Another PromptInfuser widget also exists (H) that takes the outputted artist in (B) and generates three recommended tracks.}~\label{fig:promptinfuser-interface}
  \Description{The PromptInfuser interface shown alongside a Figma UI mockup that it has made semi-functional. The UI mockup is a music search application, consisting of a search bar with decade filter. There is a recommended artist to listen to with a description, in this case, A Tribe Called Quest. At the bottom is a table of three recommended tracks with albums. On the right of this mockup is the PromptInfuser interface, which consists of a textbox, which contains the prompt, a section to specify how the prompt should be split and sent to output text elements, and finally a section to run the prompt. On the right of the PromptInfuser interface, is the full prompt that was infused into the mockup. It consists of a one-shot example of a music request for ``Calm music to work to'' which is fulfilled with the suggestion, ``Brian Eno''.}
\end{teaserfigure}

\maketitle

\section{Introduction}
\begin{figure*}
\centering
  \includegraphics[width=0.6\linewidth]{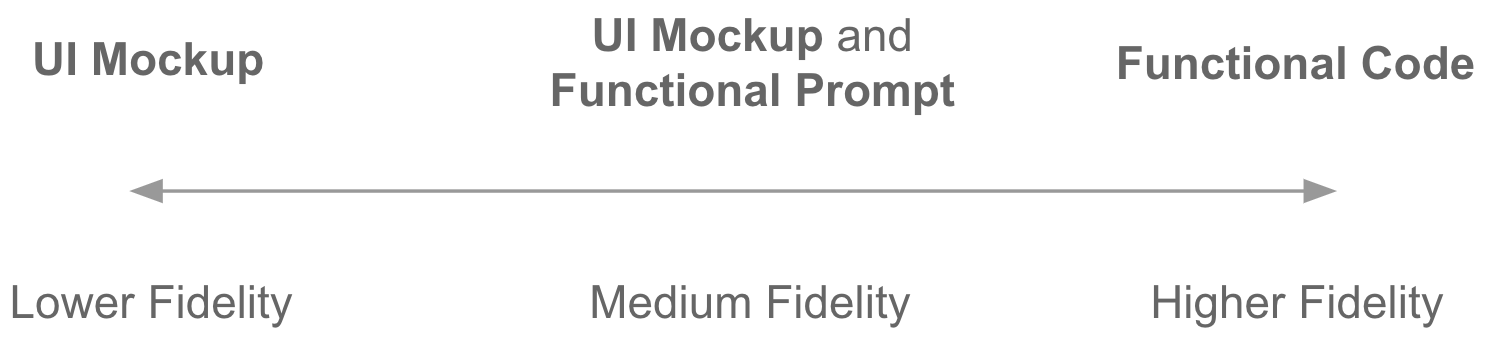}
  \caption{\textbf{Medium-fi prototypes.} Lower fidelity prototypes, like mockups, are easy to modify and experiment with but are a less realistic representation of the application, whereas higher fidelity prototypes, like functional code, provide a felt and functional experience of the application but cannot be easily altered or experimented with. \textbf{Medium fidelity prototypes} blend the ease of experimentation of LoFi prototypes and the functionality offered by HiFi prototypes; both the UI and computation (e.g. LLM prompt) are \textit{rough} and can be easily edited.}~\label{fig:prototype-spectrum}
  \Description{This figure shows a double-headed arrow, depicting the range of prototype fidelities, from lower, to medium, and finally higher fidelity. At the left, lower fidelity end, above the arrow is UI mockups. In the middle fidelity section, there is UI mockup and functional prompt. On the right, higher fidelity end, there is functional code.}
\end{figure*}

Prototyping is a core part of a designer's process.
Yet traditionally, prototyping AI applications has been difficult for designers, due to uncertainty surrounding AI’s capabilities and limitations, AI’s output complexity, and reliance on machine learning engineers \cite{ai_prototyping_hard_1, ai_prototyping_hard_2}. 
However, the introduction of large language models (LLMs) and specifically, natural language prompting \cite{few-shot-learners}, has made prototyping AI much more accessible. 
ML novices, including designers, can now quickly customize an LLM's output to fulfill a wide range of AI functionalities, from traditional ML tasks like summarization and machine translation, to futuristic applications like generating visualizations from natural language or building chatbots that can help you understand a subject.
For example, a novice can create a custom English-to-French translator with an LLM by writing a prompt with a few English and French pairs (e.g. ``English: Hello. French: Bonjour. English: Where is the bus? French: Où est le bus? English: How are you? French:''). With this prompt, the LLM will likely output the French translation: ``Comment allez-vous?''.
Recent work on \textbf{prompt-based prototyping}\cite{promptmaker} found that prompting has opened up a new class of users (e.g. non-ML experts like designers, product managers) to rapidly prototype AI functionality, better understand the capabilities and limitations of AI, and better communicate their AI product ideas to collaborators.

While prompting has dramatically reduced the barriers to prototyping AI functionality, it does so outside the context of prototyping an application with a user interface (UI).
Currently, designers are prototyping the AI functionality and UI separately; they have to imagine how their UI might interact with the prompt. 
Moreover, there are tools for rapidly prototyping UIs (e.g. Figma), and along with the rapid AI prototyping that prompting affords, there is an opportunity to combine the two, to support designers in rapidly prototyping functional AI applications.
However, relatively little is known about how combining these two processes might affect designers' workflows or the new capabilities it might afford.
This work investigates the utility of enabling designers to create low fidelity mockups that are made semi-functional with LLM prompts.
We aim to address the following central questions: \textbf{How might tightly coupling UI and AI design impact designers' workflows, and what are the benefits and drawbacks of this tight integration?}

To answer this question, we built PromptInfuser, a Figma plugin that enables designers to connect their static UI mockups to dynamically-run LLM prompts. 
Over the course of several weeks, we designed and iterated on PromptInfuser with four professional designers at a large technology company. 
From these conversations, we established how UI elements (e.g. text fields) within a mockup can be connected to the inputs and outputs of an LLM prompt. 
We refer to prompts connected to UI elements as \textbf{infused prompts}, which designers can author with PromptInfuser. 
To assign the input of an infused prompt, designers can select any number of text fields within their mockup (Figure \ref{fig:promptinfuser-interface}A) and connect their contents to input tags in the LLM prompt (Figure \ref{fig:promptinfuser-interface}D). 
To assign the infused prompt’s output, designers can either output the entirety of the prompt’s output to a \textit{single} UI element, or to \textit{multiple} UI elements (Figure \ref{fig:promptinfuser-interface}B) by splitting the model output. Finally, designers can then assign a UI element (e.g. button) in the mockup to invoke the infused prompt, so that users can interact with it naturally.

In a within-subjects user study with 14 professional designers, we compared PromptInfuser to the current design workflow of designing the UI and prompt separately (i.e. Figma and a text-based prompt-authoring interface). 
We found that, relative to existing workflows, PromptInfuser encouraged a process where designers worked on the prompt and UI in tandem, jumping back and forth from one to the other. 
Furthermore, by entering different inputs and seeing the prompt’s output dynamically appear in the UI, designers could quickly see incompatibilities between their prompt and the UI.
Finally, based on post-study survey metrics, we found that users perceived PromptInfuser to be significantly more useful for communicating the essence of a product idea, more efficient for creating the prototype, and more helpful for anticipating UI issues and potential technical constraints.

Importantly, by having both prompt and UI in mind, designers could more easily discover issues with their original problem definition, reflect on the holistic solution, and envision changes to both prompt and UI. In essence, PromptInfuser enables a space in which users can conduct \textbf{medium-fi prototyping}, by hooking up dynamic AI prototypes with low-fidelity UI prototypes.

Overall this paper makes the following contributions:
\begin{itemize}
    \item We provide a formalization of how text elements in a UI can connect to an LLM prompt. This was derived through formative studies and conversations with four professional designers.
    \item We present the design of PromptInfuser, a Figma plugin that brings together UI and AI design, by connecting dynamically-run LLM prompts to UI mockups. These infused prompts can then be dynamically invoked in the UI.
    \item We find that in a user study with 14 professional designers, participants perceived PromptInfuser to be significantly more aligned with prototyping needs: users found it more capable of creating prototypes that (1) communicate the essence of a product idea, (2) represent the envisioned artifact, and (3) anticipate both future technical constraints and UI issues. Participants also felt more efficient in creating realistic mockups with PromptInfuser, though some found it more mentally arduous to consider UI and AI simultaneously.
    \item We describe how PromptInfuser changed designers' workflows by encouraging a back-and-forth, iterative improvement of UI and prompt. We described how this workflow helped designers keep both AI and UI in mind, which in turn helped them reflect on, question, and improve the overall solution.
\end{itemize}

Taken together, these findings inform future design tools for prototyping human-AI interfaces, where UI and AI design occur in tandem.

\section{Related Work}

\subsection{Prototyping in Design}
A prototype is an approximate representation of an interactive system \cite{prototyping1}.
Designers build prototypes to test hypotheses pertaining to a design and to get feedback from users \cite{prototyping2}.
Prototyping also helps designers think through their design solution.
By building and experiencing their design first hand, as opposed to simply thinking it through, designers can reflect-in-action. That is, they can reflect on their current solution, envision changes, and better formulate what the design should be \cite{prototyping3_reflective_practitioner, reflective_physical_prototyping, reflection-in-design-2}.
Overall, prototyping supports the design process by enabling designers to test hypotheses and actively think through their design.

Because prototypes are often tailor-made to answer a particular question, they can take many different forms, with varying levels of fidelity.
For example, a low fidelity (LoFi) prototype can be a hand-drawn depiction of an interface.
These are called paper prototypes, and because they are quick and easy to create, (1) anyone can make them and (2) they offer a flexible means of testing a variety of early user flows and design ideas \cite{paper_prototype_1, paper_protoype_2, paper_prototype_3}.
LoFi prototypes are a flexible and easy means to explore the design space of solutions, and as such, they often contain very little to no functionality. Consequently, one limitation of LoFi prototypes is that they cannot be used to test accurate or ``felt experiences'' of the envisioned application.
In contrast, high fidelity (HiFi) prototypes are used to test more established designs; they are interactive and contain almost full functionality \cite{low_vs_high_fidelity, HiFi_prototypes}.
As such, relative to LoFi prototypes, HiFi prototypes take significant time and skill to produce.
PromptInfuser is a tool that enables designers to create prototypes that are simultaneously low and high fidelity.
With PromptInfuser, designers can quickly and easily create UI mockups in Figma, but also make them \textit{functional} by infusing LLM prompts, giving them a dynamic and interactive quality.
In this work, we refer to this kind of prototype as \textit{medium fidelity} (Figure \ref{fig:prototype-spectrum}) and introduce its uses, benefits, and limitations.

\subsection{Supporting Designers with Prototyping AI}

Prototyping with AI has traditionally been challenging for designers.
One major hurdle designers face when designing AI applications is that they have difficulty understanding what AI can and cannot do \cite{ai_capabilities_1, ai_capabilities_2, ai_capabilities_3, ai_capabilities_4}.
This makes brainstorming potential applications and interfaces hard to begin with.
Along with understanding AI's current capabilities, designers also have trouble with envisioning novel uses of AI. 
AI is relatively complex and can adapt over time, and this complexity hinders designers' abilities to identify many new interactions \cite{ai_capabilities_1, ai_use_cases}.
Next, even with a novel interaction in mind, it is challenging to build low-fidelity prototypes of AI, because building AI typically requires significant expertise and engineering effort. Thus, by default, designers rely on engineers to create functional prototypes\cite{scarce_ml_engineers_1, scarce_ml_engineers_2}, with the additional hurdle of needing to communicate their ideas to them in words or static mocks \cite{ml_communication_1,ml_communication_2}. To circumvent the engineering workload associated with building functional AI prototypes, designers and HCI practitioners will often create Wizard of Oz systems to simulate an interactive AI prototype, where a human might act as the AI behind the scenes \cite{woz_ai_1, woz_ai_2,woz_ai_3}.
Crucially, however, these systems lose the potential errors and idiosyncrasies that AI can often introduce.
Overall, designing with AI has been notoriously difficult because designers have a hard time understanding AI capabilities, ideating new AI uses, and quickly creating realistic AI prototypes.

To overcome these challenges, recent work has focused on understanding how and why designers would like to understand AI models and supporting these processes with interactive tools.
From interviews with UX practitioners, Liao et al. found that designers aim to understand AI models in order to eliminate risky design ideas, create transparency for users, and enable better communication and negotiation with their teams \cite{model_understanding_designers}.
To support this process of model probing and understanding, researchers recently developed \textit{fAIlureNotes}, an interactive tool that supports designers in identifying model failures across diverse user groups \cite{failure_notes}.
Going a step further, \textit{ProtoAI} embeds these model outputs into the actual interface design, assisting designers in iteratively revising their design by analyzing model breakdowns \cite{protoai}.
PromptInfuser, like \textit{ProtoAI}, embeds model outputs in the user interface design.
However, the AI incorporated in \textit{ProtoAI} is premade by engineers (e.g. an image classification model), whereas in PromptInfuser, the AI is authored and iterated on by the designers themselves (via writing and revising LLM prompts).
Accordingly, we study how the process of developing both prompt and UI in the same environment differs from their normal workflow.
Finally, our research builds on findings from preliminary work-in-progress prototypes incorporating LLMs into design environments \cite{promptinfuser_lbw}. Building on those formative designs, we provide a formalization of infused prompts; introduce a more complete plugin that lets users split LLM outputs to multiple UI outputs and invoke the LLM prompt dynamically from within the mockup; and provide findings from a 14-person within-subject study to evaluate how tightly coupling AI and UI design impacts designers' workflows and experience.

Finally, large language models have dramatically reduced the barriers to prototyping AI functionality, and in combination with tools for rapid design prototyping, like Figma, there is an opportunity to support rapid AI application prototyping.
Recent work on prompt-based prototyping found that prompting enabled non-ML experts, such as designers, to rapidly prototype AI functionality.
Through this rapid prototyping, they were better able to understand the capabilities and limitations of AI and communicate their AI product ideas to collaborators \cite{promptmaker}.
Non-ML experts can also prototype more complex AI functionalities by \textit{chaining} prompts together, where the output of one prompt is the input to another \cite{ai-chains}.
Tools such as \textit{PromptChainer} enable users to visually construct these chains of prompts and prototype sophisticated AI functionalities like music chat bots \cite{promptchainer}.
LLM prompting has made iterating over and prototyping AI functionality much faster and more accessible, but designers are still writing their prompts and creating their designs separately.
In this paper, through PromptInfuser, we study the new capabilities afforded by integrating prompts into UI mockups to rapidly create functional, low fidelity prototypes.

\subsection{Human Interaction with Generative AI and LLMs}
Generative AI, and particularly LLMs, have introduced and enabled the study of a variety of AI-backed interactive applications in HCI.
These models are being used to create prototypes that support a variety of tasks, including music composition \cite{cococo}, designing visual art \cite{opal,popblends,design_guidelines}, simulating social communities \cite{generativeagents1, generativeagents2} and writing \cite{sparks,anglekindling,wordcraft,talebrush,cells}.
Toward helping users rapidly create AI-based prototypes, there has predominantly been work on supporting the prompt writing process.
\textit{PromptIDE} automatically generates variations on a prompt for users to inspect and evaluate \cite{prompt_ide}. \textit{ScatterShot} enables users to iteratively evaluate their prompt and select effective examples to add to it \cite{scattershot}.
More recent work has proposed a framework, ``\textit{cells, generators, and lenses}'', to support users in iterating over different versions of their prompt with varying parameters \cite{cells}.
While there has been significant work on studying how users iterate over their prompts and how to improve this process, relatively little attention has been given to understanding how users develop prompts in the context of a fuller prototype with user interface.
In this work, we probe how prompt design changes when done in conjunction with UI design, from the perspective of designers.

\section{Connecting UI and Prompt}
\begin{figure*}
\centering
  \includegraphics[width=0.8\linewidth]{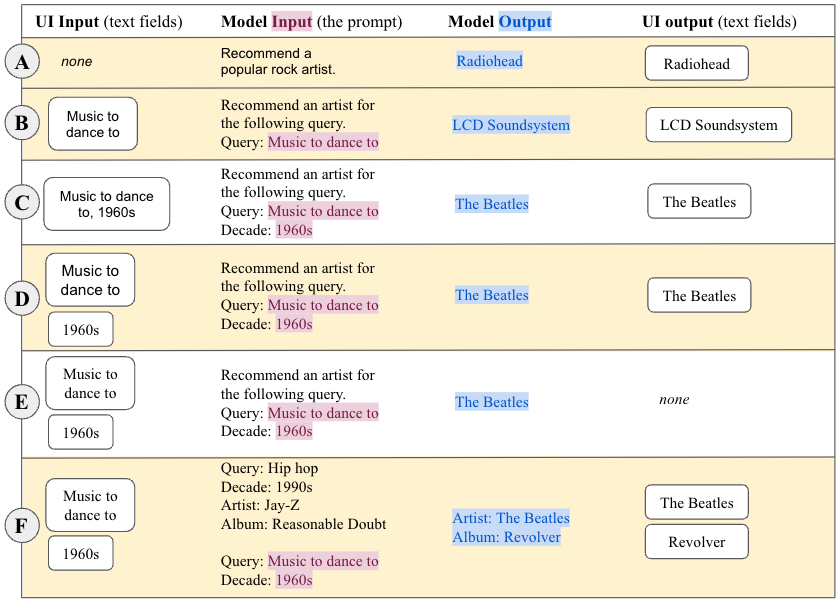}
  \caption{\textbf{Summary of how UI elements can connect to the inputs and outputs of an LLM prompt in PromptInfuser}. The rows highlighted in yellow are those we implemented for PromptInfuser. Regarding inputs, $N$ UI inputs can map to at most $N$ model inputs. That is, one UI input cannot map to more than one prompt input (C), but two UI inputs can map to two prompt inputs (B). Regarding outputs, a model output can be split to map to multiple UI outputs (F), but there needs to be at least one UI output to create an infused prompt (E). }~\label{fig:ui-prompt-mapping}
  \Description{This is a figure, depicting a table that illustrates how UI elements connect to Prompt inputs and outputs in PromptInfuser. There are four columns to this table. On the left is a column for UI Input. The next column is Model Input. The third column is Model Output. And the fourth column is UI output. This table broadly illustrates that N UI inputs map to at most N prompt inputs. For example the UI input ``Music to dance to'' maps to a single argument in a music-recommendation prompt shown in the Model Input column. This table also shows that model outputs can be all sent to one UI output or split to multiple outputs. For example, the model output ``Artist: The Beatles Album: Revolver'' is split to send the ``The Beatles'' to one UI element and ``Revolver'' to another UI element.}
\end{figure*}
Over several months, we conducted think-alouds and formative studies with four professional designers at a large technology company, and conducted multiple rounds of iteration on PromptInfuser. 
All designers had experience in designing LLM-based mockups; two of the four also had lightweight experience building high fidelity web prototypes with LLMs. 
From our discussions, we established how text elements in mockups should connect to the inputs and outputs of a prompt. Because of Figma limitations, we do not explore how other UI elements (e.g. sliders, radio buttons, drop-downs) might connect to a prompt, but we discuss the need for a general UI framework for LLMs in the discussion. 
From this point forward, UI inputs refer to input text elements.

\subsection{Inputs: Mapping UI Inputs to Model Inputs}
This section refers to the first two columns of Figure \ref{fig:ui-prompt-mapping}: \textit{UI input} and \textit{Model Input}. There are many ways UI inputs could map to the model's inputs.
The model could take no input from the UI (Figure \ref{fig:ui-prompt-mapping}A), a single UI input could map to a single model input (Figure \ref{fig:ui-prompt-mapping}B), a single UI input could map to \textit{multiple} model outputs (Figure \ref{fig:ui-prompt-mapping}C), or multiple UI inputs could map to multiple model inputs (Figure \ref{fig:ui-prompt-mapping}D).
Ultimately, we decided to implement all of these, except for a single UI input mapping to multiple model inputs (Figure \ref{fig:ui-prompt-mapping}C).
From our conversations with designers, we found that $N$ UI inputs will generally map to exactly $N$ model inputs. 
For example, in the simplest case, one UI input can map to one model input. 
Let’s say a designer is mocking up a music search application, like in Figure \ref{fig:promptinfuser-interface}, and there is a search bar where the user enters a music query like, ``music to dance to''.
This single UI input can be mapped to a single model input, which generates the artist recommendation (Figure \ref{fig:ui-prompt-mapping}B).
Similarly, two UI inputs can map to two model inputs (Figure \ref{fig:ui-prompt-mapping}D).
Semantically different inputs for an LLM prompt often come from separate text fields in the UI, such as the query and decade filter shown in Figure \ref{fig:promptinfuser-interface}A.
And so, a text field containing both of these inputs, like in Figure \ref{fig:ui-prompt-mapping}C, is generally unlikely.
Therefore, PromptInfuser enables designers to connect $N$ UI inputs to at most $N$ model inputs.


\subsection{Outputs: Mapping Model Outputs to UI Outputs}
This section refers to the second two columns of Figure \ref{fig:ui-prompt-mapping}: \textit{Model Output} and \textit{UI Output}.
Like mapping UI inputs to model inputs, there are many ways the model's output could be mapped to UI outputs.
A model output could map to: (1) a single UI output (Figure \ref{fig:ui-prompt-mapping}A), (2) multiple UI outputs (Figure \ref{fig:ui-prompt-mapping}F), (3) or no UI outputs (Figure \ref{fig:ui-prompt-mapping}E).
In this case, all of these are implemented except for mapping to no UI outputs, as the infused prompt would have no effect.
From conversations with the designers, we learned that most of their LLM-based high fidelity prototypes contained prompts that generate content for multiple elements in their UI. 
For example, in Figure \ref{fig:ui-prompt-mapping}F, the prompt is generating both an artist and a recommended album for the artist, given a music query.
The artist and album title would then be used to populate separate UI elements. 
Having a single prompt output multiple pieces of content saves time, since the model is not called multiple times to first recommend an artist and then recommend an album.
In summary, a model output must map to at least one UI output.

\section{PromptInfuser}
In this section, we introduce PromptInfuser, a Figma plugin for creating UI mockups with infused prompts. We start by explaining its design rationales and then illustrate its usage with a user walk-through. PromptInfuser enables users to create medium-fi prototypes, which include both (1) flexible and easy UI authoring, like its LoFi counterparts, as well as (2) flexible AI authoring, which make these prototypes semi-functional, similar to HiFi prototypes (illustrated further in Figure \ref{fig:prototype-spectrum}).

\subsection{Design Rationales}
In this section, we summarize a set of three design rationales for PromptInfuser we established with the same four professional designers.

\begin{itemize}
    \item[R.1] \textbf{Authentic interaction with the infused mockup}. In early prototypes of PromptInfuser, users would both (1) author the prompt and (2) run it within a pop-up window. While this made the mockup semi-functional, this interaction was not authentic. Instead, designers wanted the ability to embed the prompt within a clickable button in the UI, to enable a more realistic interaction. As explained by P1: ``It's one thing to show in our demos... `press this button' [in the pop-up] and we auto-fill a text field, and another to have a [real] submit button. If there was an availability to make something interactive...it's the most valuable piece.'' Designers imagined a major value add of a tool like PromptInfuser would be to create prototypes that closely resembled the experience of a real application.
    \item[R.2] \textbf{Easy visual inspection of Prompt and UI connections}. From feedback on early mockups of PromptInfuser, the designers wanted to easily inspect the connections between the prompt and the UI. Instead of a reference to the UI element within the prompt, designers wanted the text from that UI element to be displayed within the prompt as well as dynamically update when its content changes. At the same time, when connecting UI outputs to the prompt, we found designers wanted to be able to click on the connected outputs in the authoring panel to then highlight the corresponding UI elements in the mock. Overall, as the prompts designers wrote included more UI inputs and outputs, the visual inspection of these inspections became crucial to check if everything was set up correctly.
    \item[R.3] \textbf{Easy mapping of prompt completion to multiple UI outputs}. Two designers had experience in building web prototypes with LLMs, and the prompts they included in prototypes would often generate multiple pieces of information that they would disseminate to different elements in their UI. One pain point of this process was parsing out these separate outputs from an LLM's completion; they were not sure how to structure their prompts to produce consistent outputs to make parsing out useful information easier. Therefore, our last design rationale was to make it easy for designers to write prompts that created multiple UI outputs; splitting an LLM's completion should be relatively intuitive and not require any coding or regular expressions.
\end{itemize}

\subsection{Interface Implementation and Walk-through}
From these design rationales, we developed PromptInfuser, a Figma plugin for creating UI mockups with infused prompts.
PromptInfuser was built with Figma's plugin API \footnote{https://www.figma.com/plugin-docs/}.
Also, PromptInfuser employs a version of an LLM \footnote{anonymized for peer review} that is promptable in the same way as GPT-3, OpenAI's LLM \footnote{https://openai.com/api/}.

\subsubsection{User Walk-through}
To illustrate how PromptInfuser works, let us consider a designer named Sarah who would like to test out an idea she has for an LLM-based music recommender, like the one in Figure \ref{fig:promptinfuser-interface}.
She starts by laying out a few key elements of her design, including a search bar and decade filter (Figure \ref{fig:promptinfuser-interface}A).
She then creates a few text fields for the search results, including one for the artist and another for an artist description (Figure \ref{fig:promptinfuser-interface}B).
At this point, she'd like to write her first \textit{infused prompt}, so she pulls in an instance of the PromptInfuser plugin, which as per our first design rationale (R.1), \textit{enable authentic interaction with the mockup}, adds a clickable button onto her screen that will house and execute the prompt (Figure \ref{fig:promptinfuser-interface}C).

Her intention is to write a prompt that takes a search query and decade filter as inputs and generates a recommended artist with a corresponding description.
To implement this in PromptInfuser, she needs to write a one-shot example (Figure \ref{fig:promptinfuser-interface}D), which is a single example of the expected input and output; this example primes the model to produce completions in the same format.
In this case, her one-shot consists of a query for ``calm music to work to'' from the ``1970s'', for which she includes an example response of ``Brian Eno'' and a corresponding description.
To take in the user’s input, she then starts off the query-response pair with another \textit{Music request} and \textit{Decade} tag, but this time, she presses the pink ``Insert Input'' button above the prompt and clicks the corresponding text fields, thereby wiring up the UI input to the model input.  
As per our second design rationale (R.2), \textit{enable easy visual inspection of prompt and UI connections}, the text from the UI inputs appears in the prompt in pink blocks, which dynamically update when the designer (or end-user) changes the text in those text fields. 

With her prompt written, Sarah now needs to connect model outputs to UI elements.
Since she's generating multiple outputs (e.g. an artist and description), she selects the ``Multiple'' output button (Figure \ref{fig:promptinfuser-interface}E).
She then enters two \textit{attributes}: (1) ``Artist:'' and (2) ``Description:'', which are used to extract the corresponding information from the LLM's output.
For example, consider the LLM output shown on the bottom right of Figure \ref{fig:promptinfuser-interface}.
PromptInfuser extracts the text between each entered attribute.
That is, the LLM output is first split on ``Artist'' and the text between ``Artist'' and ``Description'' is extracted from the output (i.e. ``A Tribe Called Quest'').
Then, all the text after ``Description'' is extracted (i.e. the description text for A Tribe Called Quest).
Finally, for each attribute, she selects a UI element to display the output of that attribute.
As per our second design rationale, the current text of the UI element is shown in blue and is also clickable to highlight the element in the UI.
Thus, through this simple, attribute-based splitting, users can map their prompt outputs to multiple UI elements (R.3).

With her UI outputs set up, Sarah can now assign the prompt a stop word and temperature value, like in other prompting interfaces \cite{promptmaker} (Figure \ref{fig:promptinfuser-interface}F).
She then tests her current infused prompt by pressing the ``Run Prompt'' button, and the LLM generates ``A Tribe Called Quest'' as a recommended artist for hip hop from the 1990s (Figure \ref{fig:promptinfuser-interface}B).
Satisfied with this infused prompt, she assigns it (Figure \ref{fig:promptinfuser-interface}G) to the plugin button (Figure \ref{fig:promptinfuser-interface}C), so that a user can invoke the infused prompt by pressing ``Go''.
Now she can interact with this infused prompt authentically by altering the text in Figure \ref{fig:promptinfuser-interface}A and pressing ``Go''.
After testing this functionality a bit, she decides that her design could also provide a few tracks for the recommended artist. She then adds another PromptInfuser instance, which will take the artist text field as input and produce three songs with their corresponding albums (Figure \ref{fig:promptinfuser-interface}H).
She continues developing her UI and prompts with PromptInfuser until she has a full music recommender mock.

\section{User Study}
To understand how integrating both prompt and UI design in the same environment affects designers’ workflows, we conducted a within-subjects user study, comparing mocking up an application with PromptInfuser (i.e. a design environment where AI and UI design are integrated), versus the current designer workflow of mocking up an application with a Figma prototype and writing a prompt in a standard prompt editing interface (i.e. a separated workflow).

\begin{table*}
 \begin{tabular}{p{0.3\linewidth} | p{0.65\linewidth}}
\toprule
\textbf{Measure}          & \textbf{Statement (7-point Likert scale)}                                                                                                                      \\ \midrule
\textbf{Communicate Idea}      & Q1. With \{Workflow A/B\}, I feel like I can produce prototypes that will be useful to communicate a product idea to other designers and stakeholders.                                                  \\[0.1cm]
\textbf{Understand Feasibility}        & Q2. With \{Workflow A/B\}, I feel like I can produce prototypes that help me understand a product idea's feasibility.                                                                   \\[0.1cm]
\textbf{Experience Envisioned Artifact}     & Q3. With \{Workflow A/B\}, I feel like I can produce prototypes that give me an experience close to the envisioned artifact. \\[0.1cm]
\textbf{Anticipate UI Issues} & Q4. With \{Workflow A/B\}, I feel like I can produce prototypes that help me anticipate potential UI issues brought on by AI-powered output.                                                                                              \\[0.1cm]
\textbf{Anticipate Atypical Inputs}        & Q5. With \{Workflow A/B\}, I feel like I can produce prototypes that help me anticipate less typical user situations or user inputs.                                                           \\[0.1cm]
\textbf{Anticipate Technical Problems}      & Q6. With \{Workflow A/B\}, I feel like I can produce prototypes that help me anticipate potential technical constraints or problems that would need to be addressed in the final product.                                                           \\[0.1cm]
\textbf{Prototyping Efficiency}        & Q7. With \{Workflow A/B\}, I feel like I can efficiently create realistic mockups of my idea.                          \\[0.1cm]
\textbf{Mental Demand}   & Q8. How hard did you have to work (mentally) to produce a  prototype with \{Workflow A/B?\}                                      \\ \bottomrule
\end{tabular}
\caption{\textbf{Post-task questionnaire} filled out by participants after they created mockups with both PromptInfuser and the baseline workflow. Each statement was rated on a 7-point Likert scale.}
\Description{This table shows all the questions asked in the post-task questionnaire. There are eight measures which each have their own question. These measures include: (1) Communicate Idea, (2) Understand Feasibility, (3) Experience Envisioned Artifact, (4) Anticipate UI issues, (5) Anticipate Technical Problems, (6) Prototyping Efficiency, and (7) mental demand.}
\label{tab:questionnaire}
\end{table*}

\subsection{Procedure}
The overall outline of this study is as follows: (1) prior to the study, participants completed a 30 minute self-directed tutorial on the PromptInfuser plugin. (2) During the study, participants spent 1 hour mocking up two application briefs, one with PromptInfuser (30 minutes) and the other with Figma and the prompt editor interface (30 minutes), while thinking aloud. Condition order was counterbalanced, with application assignment per condition also balanced.
(3) After mocking these two applications, participants completed a post-study questionnaire, which compared the output prototypes of each workflow. (4) In a semi-structured interview, participants described the pros and cons of each workflow and the impact each had on their design process. The total time commitment of the study was 1 hour and 45 minutes.

The two applications the participants mocked up were both search-based, to contain a functionality that is (1) familiar to designers and (2) could be powered by relatively similar LLM prompts.
These two applications were VacationSuggester, a vacation search tool, and RecipeFinder, a recipe search tool.
We balanced which application went with which condition, so half the participants used PromptInfuser with VacationSuggester, and the other half used the baseline workflow with VacationSuggester. 
The first application participants mocked up was VacationSuggester, where users could described the vacation they’d like (e.g. “a place with great pizza”) and the application responds with two locations, a description for each, and an activity suggestion (e.g. Location: New York City. Description: NYC is famous for its thin, cheese pies. Activity: Try a margarita slice at Joe’s). The second application was RecipeFinder, where users enter ingredients and descriptors (e.g. “spicy, egg, tomato”) and the application provides two dish suggestions, each with a description, and a related recipe (e.g. Dish: Shakshuka, Description: Shakshuka is a simple dish where poached eggs are served over a flavorful and spicy tomato sauce. Related recipe: Huevos Rancheros).

To situate the task, in both conditions, participants were asked to imagine that they were creating a mockup of these early ideas to show to other designers and users the following day, to get feedback on the product concept. In the baseline condition, participants were asked to carry out their typical workflow for prototyping LLM-powered applications (found from formative studies and prior literature above), which involved: (1) creating a set of static mocks in Figma that they connect with UI-triggers in Figma's prototyping mode \footnote{https://help.figma.com/hc/en-us/articles/360040314193-Guide-to-prototyping-in-Figma}, enabling users to click through those frames as if interacting with an application, as well as (2) writing the prompt that would power this application, in a standard prompt editing interface. In the PromptInfuser condition, participants did not have the separate prompt editing interface, but instead had the plugin available to create infused prompts directly in their mockup. 

\subsection{Participants}
We recruited 14 professional designers at a large technology company (average age = 34, 3 female and 11 male, experience ranging from 5 to 30 years) via an email call for participation and word of mouth.
Eligible participants were designers who had (1) experience with Figma and (2) had written LLM prompts before; this was determined via a questionnaire sent out with the call for participation.
The interviews were conducted remotely.
Participants received a \$40 gift card for their time.

\subsection{Questionnaire}
The measures in the questionnaire, shown in Table \ref{tab:questionnaire}, are based on prior literature that has established qualities of good prototypes \cite{prototyping1, prototyping2,prototyping3_reflective_practitioner}.
For example, a prototype should be able to serve as a boundary object to easily \textit{communicate} the application to other users \cite{prototyping1} and hue relatively closely to the \textit{envisioned artifact} to elicit useful feedback \cite{prototyping2}. 
A prototype should also help its author determine the \textit{feasibility} of their idea and anticipate any potential \textit{technical constraints} that will need to be addressed in future iterations \cite{prototyping1}.
We include these fundamental aspects of prototypes to compare the outputs of the two workflows.
Finally, we also include the \textit{mental demand} measure from the NASA-TLX survey to compare mental effort required by each workflow \cite{nasa-tlx}.

\section{Findings}
\begin{figure*}
\centering
  \includegraphics[width=1\linewidth]{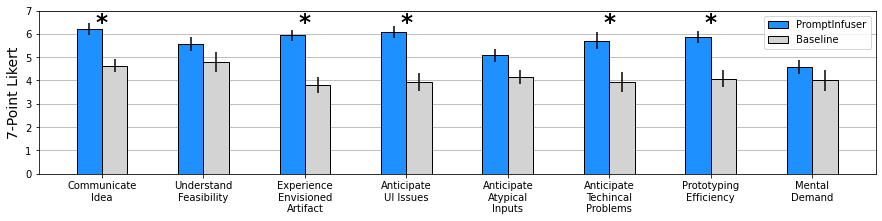}
  \caption{\textbf{Questionnaire results comparing the two conditions.} Bars are standard error and an asterisk indicates a statistically significant difference (after Bonferroni correction). }~\label{fig:quant_results}
  \Description{This is bar chart comparing PromptInfuser to the baseline condition in the user study, across all the measures from the post-task questionnaire. The y-values are from 1 to 7 for a Likert scale. In all categories, PromptInfuser scored higher than the baseline, but this difference was significant for: Communicate Idea, Experience Envisioned Artifact, Anticipate UI issues, Anticipate Technical Problems, and Prototyping Efficiency. These were indicated by asterisks.}
\end{figure*}

From the exit interviews, 13 of 14 participants \textbf{preferred} PromptInfuser to the baseline condition. Since the study was within-subjects and the questionnaire data was ordinal, we conducted eight paired sample Wilcoxon tests with full Bonferroni correction. The results are summarized in Figure \ref{fig:quant_results}. 
We found that PromptInfuser was perceived to be significantly more useful for \textbf{communicating a product idea} (Z = 0, p = .002), scoring on average 6.21 (std = 0.94), while the baseline scored 4.64 (0.97). Participants felt that being able to have collaborators enter real inputs and see results would greatly improve communication (see Section \ref{sec:social_theme}).
For similar reasons, participants felt they could create prototypes closer to the \textbf{envisioned artifact}, with PromptInfuser (Z = 2, p = .002).
Participants also felt that they could more \textbf{efficiently} create realistic mockups of their ideas, using PromptInfuser (Z = 0, p = .002).
Regarding anticipating errors, participants found PromptInfuser to be significantly more helpful toward \textbf{foreseeing potential UI issues} (Z = 0, p = .003) and \textbf{anticipating technical problems} (Z = 2, p = .005).
Participants could immediately see incompatibilities between their prompt and UI (see Section \ref{sec:layout-incompatibilities}).
Finally, no significant differences were found between the two conditions for \textbf{feasibility} (Z = 12.5, p = 0.11), \textbf{less typical inputs} (Z = 10.5, p = .07), and \textbf{mental demand} (Z = 17, p = .27).

\subsection{Two Workflows for Designing With AI}
In the following section, we provide an in depth description of the two workflows that each condition encouraged.
At a high level, while using PromptInfuser, participants \textit{went back and forth} between designing the UI and writing and revising the prompt. 
In the PromptInfuser condition, all participants (1) started by setting up a minimal UI in Figma, (2) wrote an initial prompt and connected the UI to the prompt (creating an ``infused" prompt), (3) dynamically tested the infused prompt with different inputs, and (4) based on the model output shown in the UI, revised either the prompt, UI, or both, taking them back to step 3.

Meanwhile, in the baseline condition, participants' process of designing the UI and authoring the AI \textit{tended to be disjoint} from one another, without much interleaving between the two.
Participants generally first (1) designed the UI, then (2) switched to writing and iterating on the prompt, and finally (3) copied and pasted content from the prompt to their UI, while making slight adjustments to the layout. 
Notably, the content they ultimately copied over usually represented a single run through the application in the baseline condition (or maximum three runs, if the designer was meticulous in copying copious variants), whereas with PromptInfuser, the content could be dynamically input and generated at runtime.

\subsubsection{With PromptInfuser, participants’ workflows involved a tight back and forth iteration over both the prompt and UI}
All 14 participants started by first creating a minimal version of the UI, which generally contained (1) a screen layout for the background (e.g. a vertical mobile or horizontal web layout), (2) input text field(s) for the prompt and (3) output text field(s) for the prompt's completion.
Since these text fields would eventually be overwritten by the prompt, the designers filled them with placeholders, like ``[vacation suggestion]'', instead of writing an actual example.
With this minimal UI setup, participants then started writing their prompt in the PromptInfuser plugin.
Participants either wrote the infused prompt incrementally or all at once.
Those that wrote the prompt incrementally would start with the simplest version.
For instance, in \textit{VacationSuggester}, this might be a prompt that provides two vacation locations given an input query, which users would then iterate on to add an activity suggestion for each location.
Their goal was to generally to get an infused prompt running as soon as possible.

Participants would then test their infused prompt with a wide range of inputs. To simulate what real users might enter, the designers would try both conventional inputs like ``great beaches'', as well as more eccentric ones to test the limits of the LLM, like ``adventurous jungle vacation''.
This testing phase, where designers could see a variety of completions of the LLM within the mockup, is what spurred iteration over both the UI and prompt.
Seeing the AI output in the design, participants noticed layout errors (e.g. the output is too long or short for the text element) and were inspired to think of new features or ways the user could interact with the model (see Section \ref{sec:layout-incompatibilities} and Section \ref{sec:user-changes}).
They built the UI and prompt in tandem, adjusting the layout during testing, revising the prompt to change the generated content, adding new elements to the UI, including new functionality prompt, until they were satisfied with the resulting mock.

\subsubsection{Meanwhile, in the baseline condition, participants developed the prompt and UI separately, without interweaving the two processes}
12 of 14 participants started by mocking up the UI. 
The remaining two designers started with the prompt, calling this the ``riskier'' part of the process; their ability to get useful data from the LLM was less certain than their ability to produce a great mock. 
In general, however, most participants started with the UI and often fleshed it out further than they did in the PromptInfuser condition, prior to moving onto the prompt.
One key addition they made was writing realistic content to fill in the text fields, which helped them make layout and visual hierarchy decisions.
To make the prototype feel interactive, they separated their mock into multiple frames (a typical Figma workflow that a hypothetical user could click through), which delineated the flow of the application.
Thus, in the baseline condition, designers made UI decisions based on the ``ideal'' content they had written, agnostic to the AI's actual functionality.

With a layout in place, designers then switched to the standalone prompt editor to write their prompt, with the main purpose of placing realistic content into their mock.
Some wrote multiple prompts to cover the functionality of the application (e.g. one to generate vacation suggestions and another to generate activities).
Similar to the PromptInfuser condition, designers also tested their prompts with a variety of inputs.
If they liked a particular completion, they would then use it to replace the sample content they had written in the mock.
Once they finished copying content from the prompt, however, participants did little else to update their UI, other than making a slight adjustment to the text field if the prompt output did not fit.
In the baseline, there was minimal influence of the prompt on the design and vice versa. In the following sections, we go into more depth on how AI and UI design influenced each other.

\subsection{PromptInfuser Surfaced Incompatibilities Between the UI and Prompt}
\label{sec:layout-incompatibilities}
By viewing the prompt's completions in their mock, designers could immediately see incompatibilities between the layout and prompt.
For example, P2 had set up an infused prompt for \textit{RecipeFinder}, which given an input query, generated two dishes with corresponding descriptions.
After testing this prompt with a few inputs, they noticed that (1) the two dish descriptions differed from each other in length (e.g. one might be a sentence longer than the other) and (2) the description lengths varied overall across runs (e.g. the descriptions might be two sentences one run and then three sentences in another).
Reflecting on this second inconsistency, P2 determined that she needed to ``find the right font size and layout, so it is scaleable for different types of answers'' or update the prompt to produce results that did not overflow the text fields.
These types of inconsistencies, like variations in completion length, were generally easy to miss in the Baseline prompt editor: because the prompt may produce the correct content in the correct order, the output would not seem to be an ``error''. 
However, when placed directly inside the UI, the incompatibility between the completion's length and the current layout is easily noticeable. 

Beyond surfacing these easy to miss incompatibilities, PromptInfuser also made content-based LLM errors more palpable.
For example, P14 had created an initial infused prompt for \textit{RecipeFinder}, which was meant to generate two recommended dishes for a given query.
However, even though the example in their prompt contained two recommended dishes, the prompt would often generate up to three or four recommended dishes for a query. 
This broke the mapping of the prompt's outputs to the UI (i.e. one text element would contain the first dish and the other would contain three recommended dishes).
P14 explained they would likely have ignored this error in the Baseline prompt editor, whereas PromptInfuser ``makes it more evident that there's some things that are not working well [in the prompt]...If I was running the prompt [in the Baseline], and it creates \textit{dish3} and \textit{dish4} I would just ignore it and copy and paste [the correct material]. But when hooking it up, and it wasn't working, I definitely noticed.''
In addition to this type of error, PromptInfuser also made it more palpable when the model produced (1) duplicate recommended dishes for a query or (2) failed to produce a second dish at all.
In general, connecting the prompt to the UI brought to light the prompt's content-based errors.

\subsection{PromptInfuser Supported Solution Evolution via Reflection-in-Action}
\label{sec:user-changes}
Beyond revealing surface-level incompatibilities between the layout and prompt, PromptInfuser also supported designers' deeper decision-making processes while they iterated on their design and problem definitions.
Including both the prompt and UI design in the same environment helped designers reflect-in-action.
Reflecting-in-action is essentially the process of experimenting and reflecting upon a situation as it happens, to then change one's understanding and imagine a new solution \cite{prototyping3_reflective_practitioner}. 
That is, as the designers built and interacted with both prompt and UI, they more easily kept both in mind, formed opinions on their interaction, and then better revised their solution.
In this section, we illustrate how PromptInfuser helped designers (1) adjust prompts to produce content that better fit their design, (2) ideate new UI elements to evolve the user experience, and (3) revise the prompt and design to improve user interaction. 

\subsubsection{Producing Content that Better Fit the Solution}
By showing model output in the context of a UI, PromptInfuser helped designers reflect on the kind of content they wanted the AI to produce.
While mocking up \textit{VacationSuggester}, P10 wrote an infused prompt that, for a given query, generated a suggested city, a description for the city, and two suggested activities.  
P10 then tried the input ``Burritos'' in this infused prompt, and the prompt suggested ``San Francisco'', with the following description: ``San Francisco is a renowned burrito and taco city.''
Seeing this completion in the context of the UI mock, P10 discovered aspects they could revise: ``I'm trying to think through some of the things I want to change. It's like, yeah, I could imagine adding different activities [to the prompt]...but now I also want to control the length of this [city description]. So I'll poke around to see if I can give more detail [in the prompt's one-shot example].''
P10 originally allocated much more UI space for each suggested city, and by seeing the AI output directly in the UI, they realized that the description was too short and uninformative.
P10 thus went to Wikipedia, retrieved a few more relevant facts for the city in the prompt example, and subsequently the prompt began generating longer, more informative descriptions.
Hence, viewing the AI output in the context of the user interface helped P10 more readily consider the problem from the perspective of the user. 

\subsubsection{Ideating New UI Elements}
Viewing the infused prompt (the solution as a whole) not only inspired changes to the prompt, but it also provoked revisions to the UI and application workflow.
For example, once P11 had set up the infused mock for \textit{VacationSuggester}, they inputted ``Mountain biking and beers''.
The prompt then outputted \textit{Colorado} as a suggested location, with the suggested activity: ``Try the highest mountain biking trail in the country.''
Seeing this AI output in the context of the design, P11 remarked: ``Seeing this [the suggested activity], I'm like, maybe I'd want to see an image with the activity...maybe there's a carousel...maybe there's a little link below, like `rent mountain bikes here' with a logo of the company. Like, it starts to instantly make me think of all the elements that are missing that would really make what's a just a blue rectangle right now [the placeholder `image'], a lot more compelling.'' Consolidating the UI and prompt enabled P11 to consider new user flows and related applications, such as a carousel of images to accompany the activity text, and a bike rental flow.

In turn, these newly inspired UI elements and flows often then inspired further revisions to the prompt. 
For example, like P11, P13 immediately thought of new interaction ideas as they viewed the AI's generated activity in the UI layout: ``it's so lonely and boring when it's text only. And it's easier to ignore that when you're in just a text editor [the prompt editor]...but then when you see it in a layout, I think, `That's nothing on its own. It needs an image or some supporting details'...I think that is a distinction in building something in a text-only context...versus a visual context, [where you can] recognize more quickly where your visual hierarchy should be.''
These UI inspirations in turn helped them brainstorm what modifications to the prompt structure might be needed: ``Maybe I need to be producing two different outputs here [for the vacation suggestion], like an establishment, so that I can use that as input to an image search, which is distinct from the activity description.''
Overall, PromptInfuser helped designers think through how their UI and prompt might co-evolve in tandem to produce more holistic solutions for meeting user needs.

\subsubsection{Improving User Interaction with the Prompt}
Finally, joining prompt and UI also helped designers get closer to experiencing the system from the perspective of users. 
P12 set up an infused prompt for \textit{RecipeFinder}, which generated recipe suggestions for a given free-form input.
P12 started testing this prompt with conventional inputs like ``asparagus, chicken,'' as well as less conventional inputs like ``just like mom used to make'' and ``tastes like a memory.''
While testing these inputs, they discovered that it was difficult to recall ingredients, and that this could be challenging for users. They also felt odd entering multiple ingredients into one text box. This led them to consider dropdown menus instead of textboxes, and also prompted them to use three separate ingredient inputs instead of one text field. Furthermore, from testing the more eccentric inputs, P12 found that the model outputted nonsensical dishes, such as ``a dish that tastes like a memory.''
Thus, they further constrained the input so that the three inputs specifically asked for two ingredients and one flavor profile (e.g. ``savory'').
P12 then updated the prompt structure to account for these changes, where instead of taking in a single text-field as input, it now expected two ingredient inputs and one flavor profile as input.
Reflecting on this process, P12 explained that ``prototyping back and forth between the prompt and UI can help me think through what the user input or prompt should look like.''
Thus, by interacting with both the prompt and UI in the same environment, designers considered potential user-centered issues that they may not have otherwise.

\subsection{Potential Enrichment of Design Collaboration and Communication}
\label{sec:social_theme}
Though collaboration was not the focus of this study, we discovered in post-study interviews that users felt PromptInfuser could potentially improve design collaboration and communication by (1) making idea sharing and design critique easier via consolidating prompt and UI into a single document, and (2) enabling authentic user studies in the future, where participants can interact directly with the model. We explain these two themes briefly below.

\subsubsection{Easier Sharing and Design Critique}
Six designers expressed that PromptInfuser would greatly simplify the sharing of early stage ideas.
By combining both the prompt and the mockup in a single file, designers would no longer need to share multiple links with their colleagues.
As explained by P9, ``I think it [PromptInfuser] is helpful for sharing with colleagues, like designers, for them to try it out and leave comments...we keep the conversation in one place.''
In addition to asynchronous sharing, designers also imagined that PromptInfuser could help with sharing ideas with stakeholders or conducting live design critiques.
P5 explained, ``It's better to use Workflow B [PromptInfuser] when I present in front of stakeholders, or for design critiques. They can tell me what they'd like to try and we could play with the mockup.''
An infused mockup would enable stakeholders to try their own inputs with the model and get a more realistic experience of the proposed idea.
Finally, P5 also mentioned that an infused mockup would ``definitely help explain things to engineers'' to convey how future iterations of the prompt should be constructed.
PromptInfuser may help create boundary objects \cite{boundary_object_1,boundary_object_2} for communicating ideas to cross-functional colleagues.

\subsubsection{More Authentic User Studies}
While infused mockups are useful for conveying early stage ideas to other designers, they might also be used to conduct early user studies, without requiring engineers to stand up an application.
Many participants mentioned that conducting user studies would be a core use of PromptInfuser.
P11 explained, ``It would be super cool to actually get this in the hands of users and do some sort of user testing where the buttons actually work. And it’s much more real, without having them go into [the prompt editor].''
Standard Figma prototypes do not support users entering text, much less connecting these inputs to real models, and so their flows often convey ``ideal'' usage of the application.
Seeing users enter real inputs and getting their reaction to the model's outputs is very valuable, as designers can get a gauge on (1) what inputs users might actually enter and (2) the user's authentic reaction to the application.
However, there is a limit to the authenticity PromptInfuser enables. Since the prompt has to be connected to text fields, designers cannot create the exact UI they had in mind.
P10 explained, ``There’s some limitations to...connecting the prompt to the design file, where I can’t push on all the possible things I’d want to do, like connecting a drop-down to the prompt instead of a text element.''
Therefore, there's likely room for both prototypes in a designer's workflow, where (1) PromptInfuser can be used to see how a user might interact with a model within a mockup and (2) a Figma prototype for evaluating pixel perfect designs with the requisite UI elements.

\subsection{Drawbacks of Tightly Coupling UI and AI Design}

\subsubsection{The Potential of Over-Indexing on a Particular LLM}
Although participants were able to quickly modify user interfaces in response to AI functionality, there is also some risk in over-indexing one's design to a particular version of an LLM, especially given the pace with which LLMs and machine learning models are evolving; the constraints a designer places on their design based upon an LLM's current capabilities might become obsolete later on.
As P1 put it, ``I know that they [LLM outputs] will get better quick over time...I don’t worry too much about designing around the outputs it gives me today, because I know it will be more accurate or more controllable [in the future].''
Instead, they noted it's important to create a design that is intuitive and useful for users, rather than one that compromises this in order to yield better results from the LLM.
Therefore, designers may want to lean more on using PromptInfuser to reflect on their problem definition and user mindsets, rather than altering UI purely in reaction to temporary AI deficiencies. 

\subsubsection{Frictions Introduced Through Tightly Coupling UI and AI Design}
Although a majority of users preferred PromptInfuser, some designers mentioned that because prompt and UI design involved different modes of thinking, switching rapidly from one to the other was mentally taxing. For example, P1 preferred a separation of concerns: ``I feel more free to test [the prompt] in [the prompt editor]...I'm able to structure the input and output quickly...I don't need to connect anything [to the UI]. And when I enter Figma, it's about designing the experience. So, the plugin is conflating both a little bit.''
P2 echoed these thoughts: ``I kind of feel like I'm doing a lot of things [with PromptInfuser]. While I’m working on the prototype, I feel like I can’t really focus on either of them, because I’m like, I need to make this UI perfect, but at the same time, I’m trying to figure out how to make this prompt work.''
Many participants expressed a desire to first experiment with the prompt in the prompt editor, and then be able to import that prompt to PromptInfuser.
Thus, a better workflow might allow users to first work on the prompt and UI individually, then when both reach a certain ``minimum viability,'' the two can then be combined to evolve together in PromptInfuser. We consider these issues and alternative workflows in Discussion section below (see Section \ref{sec:ideal-workflow}).

\section{Discussion}

\subsection{Toward ``Medium-Fi'' Prototypes}
Commonly used tools for prototyping UIs, like Figma, have traditionally not enabled users to include computation into their mockups. 
Previously, computation has been enabled primarily through writing code, which has generally been outside a designer's skill set.
However, natural language prompting has made authoring computation and AI functionality much faster and more accessible.
As LLMs are being incorporated into more technologies, it is becoming increasingly important to (1) design with these complex models in mind and to (2) understand how users interact with these models earlier in the design process.
There is both an opportunity and need for tools that support authoring “medium-fidelity” prototypes.
Medium-fi prototypes would be easily authored and customized, like their low fidelity counterparts, but would also include some form of dynamic computation (e.g. an LLM prompt), which is normally found in high fidelity prototypes.
A unique aspect of medium-fi prototypes is that their embedded computation, the LLM prompt, has a similar ``roughness'' to that of the LoFi UI mockup.
In a similar way a LoFi paper prototype, can be quickly revised on the spot, with components swapped out for each other, the prompt is also malleable and can be easily rewritten and assigned to different UI elements.
Future work could investigate designers' usage of medium-fi prototypes in their process, the unique needs and opportunities of medium-fi prototyping, as well as formalize the design space of medium-fi prototypes.

\subsection{Reducing Friction in Adjusting Connections Between the Prompt and UI}
From our user study, we found that adjusting connections between the UI and prompt introduced friction into the design process.
This was especially evident when designers wanted to experiment with a different type of user input in their mockup; after updating the UI, they needed to restructure their prompt to support the new user input (e.g. changing the structure of their one-shot example) and then reconnect this prompt to the UI.
A key feature of LoFi prototypes is that they are very easy to adjust and mix and match various UI components to experiment with multiple variations \cite{paper_prototype_1,paper_protoype_2}.
To allow for a similar type of flexible experimentation with infused prompts, adjusting connections between the prompt and UI should be as frictionless as possible.
Ideally, to create a new connection, the interaction would simply involve selecting a candidate prompt and then the intended UI inputs and outputs.
Then to disconnect and adjust this connection, the user could then just select new UI inputs and outputs for the prompt.
While this minimal form of interaction would certainly make adjusting connections between the prompt and UI easier, it would likely be challenging to implement.
From the selected UI inputs and outputs, the system would need to infer which parts of the prompt are inputs and connect the corresponding UI inputs.
Thus, future work might involve identifying the most intuitive and low effort way of adjusting connections between the prompt and UI.
By solving this problem, we would make flexibly experimenting with multiple infused-designs much easier; designers could quickly assign multiple prompts to alternative layouts and target a full solution faster.

\subsection{Leveraging LLMs to Make Creating and Testing Prototypes Easier}
In addition to providing computation to a LoFi prototype, LLMs can also be used to support the processes of creating and testing these infused mockups.
From the user study, 12 of 14 designers started by first creating their mockup, so one avenue of future work could be to auto-generate a prompt from this initial mockup, to help them get to an infused mockup faster. For example, designers could write example text into their UI inputs and outputs, then auto-generate an initial prompt. 
Similarly, in the opposite direction, future work could also look into support for generating a UI that would be compatible with an initial prompt.
For example, based on the freeform nature of the \textit{VacationSuggester} prompt input (e.g. ``a place with great mountain biking'') and the structured output suggestions (e.g. ``Colorado''), an LLM might recognize this as a search-like interaction and recommend a search bar. Beyond this, designers felt a key benefit of PromptInfuser was their ability to test their mockup with a variety of dynamic inputs. Given this, future research could explore the possibility of using LLMs themselves to support designers in brainstorming diverse inputs, such as through ideating user personas \cite{generativeagents1, generativeagents2} within the context of infused prompts, to better anticipate how a wide range of users might interact with their prototype.  

\subsection{Toward an Ideal Workflow for Designing AI and UI in Tandem}
\label{sec:ideal-workflow}
Finally, there remains an open question as to the ideal workflow for designing AI and UI together.
On one hand, designing both the UI and AI together helped designers discover much earlier the incompatibilities between their UI and prompt, as well as potential issues with overall problem definition. 
On the other hand, having to design both simultaneously also introduced friction, as some felt they could not fully focus on each task. 
Given this, what might an ideal workflow look like?

One potential workflow could involve first bringing both the UI and AI to an initial ``good enough'' stage in their separate environments, and then later, combining them in PromptInfuser to develop both jointly.
By starting the development of the prompt and UI separately, we could minimize context switching and let designers develop (1) a rough UI that focuses on ideal interaction and (2) a prompt that works well enough without needing to connect it to the mockup.
Then, by combining the two later, we could enable the benefits afforded by viewing the full solution and seeing how the UI and prompt interact.
One challenge, however, is establishing what is ``good enough'' for the initial prompt and UI, and helping designers determine when it might be best to start developing prompt and UI together.
For prompt writing, ``good enough'' might be adequate performance on a few diverse inputs, whereas for UI design, ``good enough'' could be an initial wireframe mockup with a user flow.
Spending too much time improving the UI and AI separately could lead to design fixation on either end: as seen in this study, the output design might not be compatible with an LLM, or vice versa.
Thus, future work can investigate better workflows for designing both AI and UI, including supporting designers' decision making processes on when to switch from designing the two separately to designing them together.

\subsection{Limitations and Future Work}

While our implementation of PromptInfuser used text-to-text LLMs, we envision future instantiations to also incorporate multimodal models, such as allowing designers to insert images into their designs.
 Though we suspect that the basic impacts on workflow would likely mirror what we discovered in this paper (e.g. easier anticipation of downstream AI-UI incompatibilities), multimodality may increase both the complexity and capabilities of the mockups one can produce with PromptInfuser, which we leave to future work.

Furthermore, 
our findings could be enriched by observing how designers incorporate PromptInfuser in a longer-term workflow.
For example, we might discover where designers incorporate PromptInfuser in their longer-term design process, such as in early user studies or design critiques.
Overall, while our current study setup yields insights on how designers develop AI and UI together in a single session, a longer-term study would yield deeper insights on how PromptInfuser might affect their holistic design process.

\section{Conclusion}
This paper introduces PromptInfuser, a plugin for infusing LLM prompts into UI mockups.
We conducted a user study with 14 professional designers, where they used PromptInfuser and a separated workflow to create mockups of two applications.
We found that PromptInfuser was perceived to be significantly more capable of creating prototypes that \textit{communicate} the essence of a product idea, better represent the \textit{envisioned artifact}, and anticipate both future \textit{technical constraints} and \textit{UI issues}. Participants also felt more efficient in creating realistic mockups with PromptInfuser.
PromptInfuser also changed designers' workflows by encouraging a back-and-forth, iterative improvement of UI and prompt.
By developing both AI and UI in tandem, designers were better able to (1) anticipate downstream AI and UI incompatibilities and (2) iterate on the total solution.
Future work can explore formalizing the design space of medium-fi prototypes, reducing friction in adjusting the connections between the prompt and UI, supporting rapid prototype creation and evaluation with LLMs, and establishing an ideal workflow for designing UI and AI jointly.
Taken together, this work informs future design tools for prototyping human-AI interfaces.



\bibliographystyle{ACM-Reference-Format}
\bibliography{main}


\begin{thebibliography}{48}


\ifx \showCODEN    \undefined \def \showCODEN     #1{\unskip}     \fi
\ifx \showDOI      \undefined \def \showDOI       #1{#1}\fi
\ifx \showISBNx    \undefined \def \showISBNx     #1{\unskip}     \fi
\ifx \showISBNxiii \undefined \def \showISBNxiii  #1{\unskip}     \fi
\ifx \showISSN     \undefined \def \showISSN      #1{\unskip}     \fi
\ifx \showLCCN     \undefined \def \showLCCN      #1{\unskip}     \fi
\ifx \shownote     \undefined \def \shownote      #1{#1}          \fi
\ifx \showarticletitle \undefined \def \showarticletitle #1{#1}   \fi
\ifx \showURL      \undefined \def \showURL       {\relax}        \fi
\providecommand\bibfield[2]{#2}
\providecommand\bibinfo[2]{#2}
\providecommand\natexlab[1]{#1}
\providecommand\showeprint[2][]{arXiv:#2}

\bibitem[Anonymous({[n.\,d.]})]%
        {promptinfuser_lbw}
\bibfield{author}{\bibinfo{person}{Anonymous}.} \bibinfo{year}{[n.\,d.]}\natexlab{}.
\newblock \showarticletitle{Anonymous for Review}.
\newblock


\bibitem[Beaudouin-Lafon and Mackay(2007)]%
        {prototyping1}
\bibfield{author}{\bibinfo{person}{Michel Beaudouin-Lafon} {and} \bibinfo{person}{Wendy~E Mackay}.} \bibinfo{year}{2007}\natexlab{}.
\newblock \showarticletitle{Prototyping tools and techniques}.
\newblock In \bibinfo{booktitle}{\emph{The human-computer interaction handbook}}. \bibinfo{publisher}{CRC Press}, \bibinfo{pages}{1043--1066}.
\newblock


\bibitem[Brown et~al\mbox{.}(2020)]%
        {few-shot-learners}
\bibfield{author}{\bibinfo{person}{Tom Brown}, \bibinfo{person}{Benjamin Mann}, \bibinfo{person}{Nick Ryder}, \bibinfo{person}{Melanie Subbiah}, \bibinfo{person}{Jared~D Kaplan}, \bibinfo{person}{Prafulla Dhariwal}, \bibinfo{person}{Arvind Neelakantan}, \bibinfo{person}{Pranav Shyam}, \bibinfo{person}{Girish Sastry}, \bibinfo{person}{Amanda Askell}, \bibinfo{person}{Sandhini Agarwal}, \bibinfo{person}{Ariel Herbert-Voss}, \bibinfo{person}{Gretchen Krueger}, \bibinfo{person}{Tom Henighan}, \bibinfo{person}{Rewon Child}, \bibinfo{person}{Aditya Ramesh}, \bibinfo{person}{Daniel Ziegler}, \bibinfo{person}{Jeffrey Wu}, \bibinfo{person}{Clemens Winter}, \bibinfo{person}{Chris Hesse}, \bibinfo{person}{Mark Chen}, \bibinfo{person}{Eric Sigler}, \bibinfo{person}{Mateusz Litwin}, \bibinfo{person}{Scott Gray}, \bibinfo{person}{Benjamin Chess}, \bibinfo{person}{Jack Clark}, \bibinfo{person}{Christopher Berner}, \bibinfo{person}{Sam McCandlish}, \bibinfo{person}{Alec Radford}, \bibinfo{person}{Ilya Sutskever}, {and}
  \bibinfo{person}{Dario Amodei}.} \bibinfo{year}{2020}\natexlab{}.
\newblock \showarticletitle{Language Models are Few-Shot Learners}. In \bibinfo{booktitle}{\emph{Advances in Neural Information Processing Systems}}, \bibfield{editor}{\bibinfo{person}{H.~Larochelle}, \bibinfo{person}{M.~Ranzato}, \bibinfo{person}{R.~Hadsell}, \bibinfo{person}{M.F. Balcan}, {and} \bibinfo{person}{H.~Lin}} (Eds.), Vol.~\bibinfo{volume}{33}. \bibinfo{publisher}{Curran Associates, Inc.}, \bibinfo{pages}{1877--1901}.
\newblock
\urldef\tempurl%
\url{https://proceedings.neurips.cc/paper_files/paper/2020/file/1457c0d6bfcb4967418bfb8ac142f64a-Paper.pdf}
\showURL{%
\tempurl}


\bibitem[Chung et~al\mbox{.}(2022)]%
        {talebrush}
\bibfield{author}{\bibinfo{person}{John Joon~Young Chung}, \bibinfo{person}{Wooseok Kim}, \bibinfo{person}{Kang~Min Yoo}, \bibinfo{person}{Hwaran Lee}, \bibinfo{person}{Eytan Adar}, {and} \bibinfo{person}{Minsuk Chang}.} \bibinfo{year}{2022}\natexlab{}.
\newblock \showarticletitle{TaleBrush: Sketching Stories with Generative Pretrained Language Models}. In \bibinfo{booktitle}{\emph{Proceedings of the 2022 CHI Conference on Human Factors in Computing Systems}} (New Orleans, LA, USA) \emph{(\bibinfo{series}{CHI '22})}. \bibinfo{publisher}{Association for Computing Machinery}, \bibinfo{address}{New York, NY, USA}, Article \bibinfo{articleno}{209}, \bibinfo{numpages}{19}~pages.
\newblock
\showISBNx{9781450391573}
\urldef\tempurl%
\url{https://doi.org/10.1145/3491102.3501819}
\showDOI{\tempurl}


\bibitem[Cranshaw et~al\mbox{.}(2017)]%
        {woz_ai_1}
\bibfield{author}{\bibinfo{person}{Justin Cranshaw}, \bibinfo{person}{Emad Elwany}, \bibinfo{person}{Todd Newman}, \bibinfo{person}{Rafal Kocielnik}, \bibinfo{person}{Bowen Yu}, \bibinfo{person}{Sandeep Soni}, \bibinfo{person}{Jaime Teevan}, {and} \bibinfo{person}{Andr\'{e}s Monroy-Hern\'{a}ndez}.} \bibinfo{year}{2017}\natexlab{}.
\newblock \showarticletitle{Calendar.Help: Designing a Workflow-Based Scheduling Agent with Humans in the Loop}. In \bibinfo{booktitle}{\emph{Proceedings of the 2017 CHI Conference on Human Factors in Computing Systems}} (Denver, Colorado, USA) \emph{(\bibinfo{series}{CHI '17})}. \bibinfo{publisher}{Association for Computing Machinery}, \bibinfo{address}{New York, NY, USA}, \bibinfo{pages}{2382–2393}.
\newblock
\showISBNx{9781450346559}
\urldef\tempurl%
\url{https://doi.org/10.1145/3025453.3025780}
\showDOI{\tempurl}


\bibitem[Dove et~al\mbox{.}(2017)]%
        {ai_capabilities_1}
\bibfield{author}{\bibinfo{person}{Graham Dove}, \bibinfo{person}{Kim Halskov}, \bibinfo{person}{Jodi Forlizzi}, {and} \bibinfo{person}{John Zimmerman}.} \bibinfo{year}{2017}\natexlab{}.
\newblock \showarticletitle{UX Design Innovation: Challenges for Working with Machine Learning as a Design Material}. In \bibinfo{booktitle}{\emph{Proceedings of the 2017 CHI Conference on Human Factors in Computing Systems}} (Denver, Colorado, USA) \emph{(\bibinfo{series}{CHI '17})}. \bibinfo{publisher}{Association for Computing Machinery}, \bibinfo{address}{New York, NY, USA}, \bibinfo{pages}{278–288}.
\newblock
\showISBNx{9781450346559}
\urldef\tempurl%
\url{https://doi.org/10.1145/3025453.3025739}
\showDOI{\tempurl}


\bibitem[Gero et~al\mbox{.}(2022)]%
        {sparks}
\bibfield{author}{\bibinfo{person}{Katy~Ilonka Gero}, \bibinfo{person}{Vivian Liu}, {and} \bibinfo{person}{Lydia Chilton}.} \bibinfo{year}{2022}\natexlab{}.
\newblock \showarticletitle{Sparks: Inspiration for Science Writing Using Language Models}. In \bibinfo{booktitle}{\emph{Proceedings of the 2022 ACM Designing Interactive Systems Conference}} (Virtual Event, Australia) \emph{(\bibinfo{series}{DIS '22})}. \bibinfo{publisher}{Association for Computing Machinery}, \bibinfo{address}{New York, NY, USA}, \bibinfo{pages}{1002–1019}.
\newblock
\showISBNx{9781450393584}
\urldef\tempurl%
\url{https://doi.org/10.1145/3532106.3533533}
\showDOI{\tempurl}


\bibitem[Girardin and Lathia(2017)]%
        {scarce_ml_engineers_2}
\bibfield{author}{\bibinfo{person}{Fabien Girardin} {and} \bibinfo{person}{Neal Lathia}.} \bibinfo{year}{2017}\natexlab{}.
\newblock \showarticletitle{When user experience designers partner with data scientists}. In \bibinfo{booktitle}{\emph{2017 AAAI Spring Symposium Series}}.
\newblock


\bibitem[Hart and Staveland(1988)]%
        {nasa-tlx}
\bibfield{author}{\bibinfo{person}{Sandra~G. Hart} {and} \bibinfo{person}{Lowell~E. Staveland}.} \bibinfo{year}{1988}\natexlab{}.
\newblock \showarticletitle{Development of NASA-TLX (Task Load Index): Results of Empirical and Theoretical Research}.
\newblock In \bibinfo{booktitle}{\emph{Human Mental Workload}}, \bibfield{editor}{\bibinfo{person}{Peter~A. Hancock} {and} \bibinfo{person}{Najmedin Meshkati}} (Eds.). \bibinfo{series}{Advances in Psychology}, Vol.~\bibinfo{volume}{52}. \bibinfo{publisher}{North-Holland}, \bibinfo{pages}{139--183}.
\newblock
\showISSN{0166-4115}
\urldef\tempurl%
\url{https://doi.org/10.1016/S0166-4115(08)62386-9}
\showDOI{\tempurl}


\bibitem[Hartmann et~al\mbox{.}(2006)]%
        {reflective_physical_prototyping}
\bibfield{author}{\bibinfo{person}{Bj\"{o}rn Hartmann}, \bibinfo{person}{Scott~R. Klemmer}, \bibinfo{person}{Michael Bernstein}, \bibinfo{person}{Leith Abdulla}, \bibinfo{person}{Brandon Burr}, \bibinfo{person}{Avi Robinson-Mosher}, {and} \bibinfo{person}{Jennifer Gee}.} \bibinfo{year}{2006}\natexlab{}.
\newblock \showarticletitle{Reflective Physical Prototyping through Integrated Design, Test, and Analysis}. In \bibinfo{booktitle}{\emph{Proceedings of the 19th Annual ACM Symposium on User Interface Software and Technology}} (Montreux, Switzerland) \emph{(\bibinfo{series}{UIST '06})}. \bibinfo{publisher}{Association for Computing Machinery}, \bibinfo{address}{New York, NY, USA}, \bibinfo{pages}{299–308}.
\newblock
\showISBNx{1595933131}
\urldef\tempurl%
\url{https://doi.org/10.1145/1166253.1166300}
\showDOI{\tempurl}


\bibitem[Holmquist(2017)]%
        {ai_capabilities_2}
\bibfield{author}{\bibinfo{person}{Lars~Erik Holmquist}.} \bibinfo{year}{2017}\natexlab{}.
\newblock \showarticletitle{Intelligence on Tap: Artificial Intelligence as a New Design Material}.
\newblock \bibinfo{journal}{\emph{Interactions}} \bibinfo{volume}{24}, \bibinfo{number}{4} (\bibinfo{date}{jun} \bibinfo{year}{2017}), \bibinfo{pages}{28–33}.
\newblock
\showISSN{1072-5520}
\urldef\tempurl%
\url{https://doi.org/10.1145/3085571}
\showDOI{\tempurl}


\bibitem[Hong et~al\mbox{.}(2021)]%
        {ai_prototyping_hard_1}
\bibfield{author}{\bibinfo{person}{Matthew~K. Hong}, \bibinfo{person}{Adam Fourney}, \bibinfo{person}{Derek DeBellis}, {and} \bibinfo{person}{Saleema Amershi}.} \bibinfo{year}{2021}\natexlab{}.
\newblock \showarticletitle{Planning for Natural Language Failures with the AI Playbook}. In \bibinfo{booktitle}{\emph{Proceedings of the 2021 CHI Conference on Human Factors in Computing Systems}} (Yokohama, Japan) \emph{(\bibinfo{series}{CHI '21})}. \bibinfo{publisher}{Association for Computing Machinery}, \bibinfo{address}{New York, NY, USA}, Article \bibinfo{articleno}{386}, \bibinfo{numpages}{11}~pages.
\newblock
\showISBNx{9781450380966}
\urldef\tempurl%
\url{https://doi.org/10.1145/3411764.3445735}
\showDOI{\tempurl}


\bibitem[Jiang et~al\mbox{.}(2022)]%
        {promptmaker}
\bibfield{author}{\bibinfo{person}{Ellen Jiang}, \bibinfo{person}{Kristen Olson}, \bibinfo{person}{Edwin Toh}, \bibinfo{person}{Alejandra Molina}, \bibinfo{person}{Aaron Donsbach}, \bibinfo{person}{Michael Terry}, {and} \bibinfo{person}{Carrie~J Cai}.} \bibinfo{year}{2022}\natexlab{}.
\newblock \showarticletitle{PromptMaker: Prompt-Based Prototyping with Large Language Models}. In \bibinfo{booktitle}{\emph{Extended Abstracts of the 2022 CHI Conference on Human Factors in Computing Systems}} (New Orleans, LA, USA) \emph{(\bibinfo{series}{CHI EA '22})}. \bibinfo{publisher}{Association for Computing Machinery}, \bibinfo{address}{New York, NY, USA}, Article \bibinfo{articleno}{35}, \bibinfo{numpages}{8}~pages.
\newblock
\showISBNx{9781450391566}
\urldef\tempurl%
\url{https://doi.org/10.1145/3491101.3503564}
\showDOI{\tempurl}


\bibitem[John et~al\mbox{.}(2004)]%
        {boundary_object_1}
\bibfield{author}{\bibinfo{person}{Bonnie~E. John}, \bibinfo{person}{Len Bass}, \bibinfo{person}{Rick Kazman}, {and} \bibinfo{person}{Eugene Chen}.} \bibinfo{year}{2004}\natexlab{}.
\newblock \showarticletitle{Identifying Gaps between HCI, Software Engineering, and Design, and Boundary Objects to Bridge Them}. In \bibinfo{booktitle}{\emph{CHI '04 Extended Abstracts on Human Factors in Computing Systems}} (Vienna, Austria) \emph{(\bibinfo{series}{CHI EA '04})}. \bibinfo{publisher}{Association for Computing Machinery}, \bibinfo{address}{New York, NY, USA}, \bibinfo{pages}{1723–1724}.
\newblock
\showISBNx{1581137036}
\urldef\tempurl%
\url{https://doi.org/10.1145/985921.986201}
\showDOI{\tempurl}


\bibitem[Kayacik et~al\mbox{.}(2019)]%
        {ml_communication_1}
\bibfield{author}{\bibinfo{person}{Claire Kayacik}, \bibinfo{person}{Sherol Chen}, \bibinfo{person}{Signe Noerly}, \bibinfo{person}{Jess Holbrook}, \bibinfo{person}{Adam Roberts}, {and} \bibinfo{person}{Douglas Eck}.} \bibinfo{year}{2019}\natexlab{}.
\newblock \showarticletitle{Identifying the Intersections: User Experience + Research Scientist Collaboration in a Generative Machine Learning Interface}. In \bibinfo{booktitle}{\emph{Extended Abstracts of the 2019 CHI Conference on Human Factors in Computing Systems}} (Glasgow, Scotland Uk) \emph{(\bibinfo{series}{CHI EA '19})}. \bibinfo{publisher}{Association for Computing Machinery}, \bibinfo{address}{New York, NY, USA}, \bibinfo{pages}{1–8}.
\newblock
\showISBNx{9781450359719}
\urldef\tempurl%
\url{https://doi.org/10.1145/3290607.3299059}
\showDOI{\tempurl}


\bibitem[Kim et~al\mbox{.}(2023)]%
        {cells}
\bibfield{author}{\bibinfo{person}{Tae~Soo Kim}, \bibinfo{person}{Yoonjoo Lee}, \bibinfo{person}{Minsuk Chang}, {and} \bibinfo{person}{Juho Kim}.} \bibinfo{year}{2023}\natexlab{}.
\newblock \showarticletitle{Cells, Generators, and Lenses: Design Framework for Object-Oriented Interaction with Large Language Models}.
\newblock  (\bibinfo{year}{2023}).
\newblock
\showISBNx{979840070132}
\urldef\tempurl%
\url{https://doi.org/10.1145/3586183.3606833}
\showDOI{\tempurl}


\bibitem[Klemmer et~al\mbox{.}(2006)]%
        {reflection-in-design-2}
\bibfield{author}{\bibinfo{person}{Scott~R. Klemmer}, \bibinfo{person}{Bj\"{o}rn Hartmann}, {and} \bibinfo{person}{Leila Takayama}.} \bibinfo{year}{2006}\natexlab{}.
\newblock \showarticletitle{How Bodies Matter: Five Themes for Interaction Design}. In \bibinfo{booktitle}{\emph{Proceedings of the 6th Conference on Designing Interactive Systems}} (University Park, PA, USA) \emph{(\bibinfo{series}{DIS '06})}. \bibinfo{publisher}{Association for Computing Machinery}, \bibinfo{address}{New York, NY, USA}, \bibinfo{pages}{140–149}.
\newblock
\showISBNx{1595933670}
\urldef\tempurl%
\url{https://doi.org/10.1145/1142405.1142429}
\showDOI{\tempurl}


\bibitem[Klemmer et~al\mbox{.}(2000)]%
        {woz_ai_2}
\bibfield{author}{\bibinfo{person}{Scott~R. Klemmer}, \bibinfo{person}{Anoop~K. Sinha}, \bibinfo{person}{Jack Chen}, \bibinfo{person}{James~A. Landay}, \bibinfo{person}{Nadeem Aboobaker}, {and} \bibinfo{person}{Annie Wang}.} \bibinfo{year}{2000}\natexlab{}.
\newblock \showarticletitle{Suede: A Wizard of Oz Prototyping Tool for Speech User Interfaces}. In \bibinfo{booktitle}{\emph{Proceedings of the 13th Annual ACM Symposium on User Interface Software and Technology}} (San Diego, California, USA) \emph{(\bibinfo{series}{UIST '00})}. \bibinfo{publisher}{Association for Computing Machinery}, \bibinfo{address}{New York, NY, USA}, \bibinfo{pages}{1–10}.
\newblock
\showISBNx{1581132123}
\urldef\tempurl%
\url{https://doi.org/10.1145/354401.354406}
\showDOI{\tempurl}


\bibitem[Lee(2007)]%
        {boundary_object_2}
\bibfield{author}{\bibinfo{person}{Charlotte~P Lee}.} \bibinfo{year}{2007}\natexlab{}.
\newblock \showarticletitle{Boundary negotiating artifacts: Unbinding the routine of boundary objects and embracing chaos in collaborative work}.
\newblock \bibinfo{journal}{\emph{Computer Supported Cooperative Work (CSCW)}}  \bibinfo{volume}{16} (\bibinfo{year}{2007}), \bibinfo{pages}{307--339}.
\newblock


\bibitem[Liao et~al\mbox{.}(2023)]%
        {model_understanding_designers}
\bibfield{author}{\bibinfo{person}{Q.~Vera Liao}, \bibinfo{person}{Hariharan Subramonyam}, \bibinfo{person}{Jennifer Wang}, {and} \bibinfo{person}{Jennifer Wortman~Vaughan}.} \bibinfo{year}{2023}\natexlab{}.
\newblock \showarticletitle{Designerly Understanding: Information Needs for Model Transparency to Support Design Ideation for AI-Powered User Experience}. In \bibinfo{booktitle}{\emph{Proceedings of the 2023 CHI Conference on Human Factors in Computing Systems}} (Hamburg, Germany) \emph{(\bibinfo{series}{CHI '23})}. \bibinfo{publisher}{Association for Computing Machinery}, \bibinfo{address}{New York, NY, USA}, Article \bibinfo{articleno}{9}, \bibinfo{numpages}{21}~pages.
\newblock
\showISBNx{9781450394215}
\urldef\tempurl%
\url{https://doi.org/10.1145/3544548.3580652}
\showDOI{\tempurl}


\bibitem[Lim et~al\mbox{.}(2008)]%
        {prototyping2}
\bibfield{author}{\bibinfo{person}{Youn-Kyung Lim}, \bibinfo{person}{Erik Stolterman}, {and} \bibinfo{person}{Josh Tenenberg}.} \bibinfo{year}{2008}\natexlab{}.
\newblock \showarticletitle{The Anatomy of Prototypes: Prototypes as Filters, Prototypes as Manifestations of Design Ideas}.
\newblock \bibinfo{journal}{\emph{ACM Trans. Comput.-Hum. Interact.}} \bibinfo{volume}{15}, \bibinfo{number}{2}, Article \bibinfo{articleno}{7} (\bibinfo{date}{jul} \bibinfo{year}{2008}), \bibinfo{numpages}{27}~pages.
\newblock
\showISSN{1073-0516}
\urldef\tempurl%
\url{https://doi.org/10.1145/1375761.1375762}
\showDOI{\tempurl}


\bibitem[Liu and Chilton(2022)]%
        {design_guidelines}
\bibfield{author}{\bibinfo{person}{Vivian Liu} {and} \bibinfo{person}{Lydia~B Chilton}.} \bibinfo{year}{2022}\natexlab{}.
\newblock \showarticletitle{Design Guidelines for Prompt Engineering Text-to-Image Generative Models}. In \bibinfo{booktitle}{\emph{Proceedings of the 2022 CHI Conference on Human Factors in Computing Systems}} (New Orleans, LA, USA) \emph{(\bibinfo{series}{CHI '22})}. \bibinfo{publisher}{Association for Computing Machinery}, \bibinfo{address}{New York, NY, USA}, Article \bibinfo{articleno}{384}, \bibinfo{numpages}{23}~pages.
\newblock
\showISBNx{9781450391573}
\urldef\tempurl%
\url{https://doi.org/10.1145/3491102.3501825}
\showDOI{\tempurl}


\bibitem[Liu et~al\mbox{.}(2022)]%
        {opal}
\bibfield{author}{\bibinfo{person}{Vivian Liu}, \bibinfo{person}{Han Qiao}, {and} \bibinfo{person}{Lydia Chilton}.} \bibinfo{year}{2022}\natexlab{}.
\newblock \showarticletitle{Opal: Multimodal Image Generation for News Illustration}. In \bibinfo{booktitle}{\emph{Proceedings of the 35th Annual ACM Symposium on User Interface Software and Technology}} (Bend, OR, USA) \emph{(\bibinfo{series}{UIST '22})}. \bibinfo{publisher}{Association for Computing Machinery}, \bibinfo{address}{New York, NY, USA}, Article \bibinfo{articleno}{73}, \bibinfo{numpages}{17}~pages.
\newblock
\showISBNx{9781450393201}
\urldef\tempurl%
\url{https://doi.org/10.1145/3526113.3545621}
\showDOI{\tempurl}


\bibitem[Louie et~al\mbox{.}(2020)]%
        {cococo}
\bibfield{author}{\bibinfo{person}{Ryan Louie}, \bibinfo{person}{Andy Coenen}, \bibinfo{person}{Cheng~Zhi Huang}, \bibinfo{person}{Michael Terry}, {and} \bibinfo{person}{Carrie~J. Cai}.} \bibinfo{year}{2020}\natexlab{}.
\newblock \showarticletitle{Novice-AI Music Co-Creation via AI-Steering Tools for Deep Generative Models}. In \bibinfo{booktitle}{\emph{Proceedings of the 2020 CHI Conference on Human Factors in Computing Systems}} (Honolulu, HI, USA) \emph{(\bibinfo{series}{CHI '20})}. \bibinfo{publisher}{Association for Computing Machinery}, \bibinfo{address}{New York, NY, USA}, \bibinfo{pages}{1–13}.
\newblock
\showISBNx{9781450367080}
\urldef\tempurl%
\url{https://doi.org/10.1145/3313831.3376739}
\showDOI{\tempurl}


\bibitem[Moore et~al\mbox{.}(2023)]%
        {failure_notes}
\bibfield{author}{\bibinfo{person}{Steven Moore}, \bibinfo{person}{Q.~Vera Liao}, {and} \bibinfo{person}{Hariharan Subramonyam}.} \bibinfo{year}{2023}\natexlab{}.
\newblock \showarticletitle{FAIlureNotes: Supporting Designers in Understanding the Limits of AI Models for Computer Vision Tasks}. In \bibinfo{booktitle}{\emph{Proceedings of the 2023 CHI Conference on Human Factors in Computing Systems}} (Hamburg, Germany) \emph{(\bibinfo{series}{CHI '23})}. \bibinfo{publisher}{Association for Computing Machinery}, \bibinfo{address}{New York, NY, USA}, Article \bibinfo{articleno}{10}, \bibinfo{numpages}{19}~pages.
\newblock
\showISBNx{9781450394215}
\urldef\tempurl%
\url{https://doi.org/10.1145/3544548.3581242}
\showDOI{\tempurl}


\bibitem[Olivier et~al\mbox{.}(2009)]%
        {HiFi_prototypes}
\bibfield{author}{\bibinfo{person}{Patrick Olivier}, \bibinfo{person}{Guangyou Xu}, \bibinfo{person}{Andrew Monk}, {and} \bibinfo{person}{Jesse Hoey}.} \bibinfo{year}{2009}\natexlab{}.
\newblock \showarticletitle{Ambient Kitchen: Designing Situated Services Using a High Fidelity Prototyping Environment}. In \bibinfo{booktitle}{\emph{Proceedings of the 2nd International Conference on PErvasive Technologies Related to Assistive Environments}} (Corfu, Greece) \emph{(\bibinfo{series}{PETRA '09})}. \bibinfo{publisher}{Association for Computing Machinery}, \bibinfo{address}{New York, NY, USA}, Article \bibinfo{articleno}{47}, \bibinfo{numpages}{7}~pages.
\newblock
\showISBNx{9781605584096}
\urldef\tempurl%
\url{https://doi.org/10.1145/1579114.1579161}
\showDOI{\tempurl}


\bibitem[Park et~al\mbox{.}(2023)]%
        {generativeagents2}
\bibfield{author}{\bibinfo{person}{Joon~Sung Park}, \bibinfo{person}{Joseph~C. O'Brien}, \bibinfo{person}{Carrie~J. Cai}, \bibinfo{person}{Meredith~Ringel Morris}, \bibinfo{person}{Percy Liang}, {and} \bibinfo{person}{Michael~S. Bernstein}.} \bibinfo{year}{2023}\natexlab{}.
\newblock \bibinfo{title}{Generative Agents: Interactive Simulacra of Human Behavior}.
\newblock
\newblock
\showeprint[arxiv]{2304.03442}~[cs.HC]


\bibitem[Park et~al\mbox{.}(2022)]%
        {generativeagents1}
\bibfield{author}{\bibinfo{person}{Joon~Sung Park}, \bibinfo{person}{Lindsay Popowski}, \bibinfo{person}{Carrie Cai}, \bibinfo{person}{Meredith~Ringel Morris}, \bibinfo{person}{Percy Liang}, {and} \bibinfo{person}{Michael~S. Bernstein}.} \bibinfo{year}{2022}\natexlab{}.
\newblock \showarticletitle{Social Simulacra: Creating Populated Prototypes for Social Computing Systems}. In \bibinfo{booktitle}{\emph{Proceedings of the 35th Annual ACM Symposium on User Interface Software and Technology}} (Bend, OR, USA) \emph{(\bibinfo{series}{UIST '22})}. \bibinfo{publisher}{Association for Computing Machinery}, \bibinfo{address}{New York, NY, USA}, Article \bibinfo{articleno}{74}, \bibinfo{numpages}{18}~pages.
\newblock
\showISBNx{9781450393201}
\urldef\tempurl%
\url{https://doi.org/10.1145/3526113.3545616}
\showDOI{\tempurl}


\bibitem[Petridis et~al\mbox{.}(2023)]%
        {anglekindling}
\bibfield{author}{\bibinfo{person}{Savvas Petridis}, \bibinfo{person}{Nicholas Diakopoulos}, \bibinfo{person}{Kevin Crowston}, \bibinfo{person}{Mark Hansen}, \bibinfo{person}{Keren Henderson}, \bibinfo{person}{Stan Jastrzebski}, \bibinfo{person}{Jeffrey~V Nickerson}, {and} \bibinfo{person}{Lydia~B Chilton}.} \bibinfo{year}{2023}\natexlab{}.
\newblock \showarticletitle{AngleKindling: Supporting Journalistic Angle Ideation with Large Language Models}. In \bibinfo{booktitle}{\emph{Proceedings of the 2023 CHI Conference on Human Factors in Computing Systems}} (Hamburg, Germany) \emph{(\bibinfo{series}{CHI '23})}. \bibinfo{publisher}{Association for Computing Machinery}, \bibinfo{address}{New York, NY, USA}, Article \bibinfo{articleno}{225}, \bibinfo{numpages}{16}~pages.
\newblock
\showISBNx{9781450394215}
\urldef\tempurl%
\url{https://doi.org/10.1145/3544548.3580907}
\showDOI{\tempurl}


\bibitem[Rettig(1994)]%
        {paper_prototype_1}
\bibfield{author}{\bibinfo{person}{Marc Rettig}.} \bibinfo{year}{1994}\natexlab{}.
\newblock \showarticletitle{Prototyping for Tiny Fingers}.
\newblock \bibinfo{journal}{\emph{Commun. ACM}} \bibinfo{volume}{37}, \bibinfo{number}{4} (\bibinfo{date}{apr} \bibinfo{year}{1994}), \bibinfo{pages}{21–27}.
\newblock
\showISSN{0001-0782}
\urldef\tempurl%
\url{https://doi.org/10.1145/175276.175288}
\showDOI{\tempurl}


\bibitem[Riek(2012)]%
        {woz_ai_3}
\bibfield{author}{\bibinfo{person}{Laurel~D. Riek}.} \bibinfo{year}{2012}\natexlab{}.
\newblock \showarticletitle{Wizard of Oz Studies in HRI: A Systematic Review and New Reporting Guidelines}.
\newblock \bibinfo{journal}{\emph{J. Hum.-Robot Interact.}} \bibinfo{volume}{1}, \bibinfo{number}{1} (\bibinfo{date}{jul} \bibinfo{year}{2012}), \bibinfo{pages}{119–136}.
\newblock
\urldef\tempurl%
\url{https://doi.org/10.5898/JHRI.1.1.Riek}
\showDOI{\tempurl}


\bibitem[Rudd et~al\mbox{.}(1996)]%
        {low_vs_high_fidelity}
\bibfield{author}{\bibinfo{person}{Jim Rudd}, \bibinfo{person}{Ken Stern}, {and} \bibinfo{person}{Scott Isensee}.} \bibinfo{year}{1996}\natexlab{}.
\newblock \showarticletitle{Low vs. High-Fidelity Prototyping Debate}.
\newblock \bibinfo{journal}{\emph{Interactions}} \bibinfo{volume}{3}, \bibinfo{number}{1} (\bibinfo{date}{jan} \bibinfo{year}{1996}), \bibinfo{pages}{76–85}.
\newblock
\showISSN{1072-5520}
\urldef\tempurl%
\url{https://doi.org/10.1145/223500.223514}
\showDOI{\tempurl}


\bibitem[Sch{\"o}n(2017)]%
        {prototyping3_reflective_practitioner}
\bibfield{author}{\bibinfo{person}{Donald~A Sch{\"o}n}.} \bibinfo{year}{2017}\natexlab{}.
\newblock \bibinfo{booktitle}{\emph{The reflective practitioner: How professionals think in action}}.
\newblock \bibinfo{publisher}{Routledge}.
\newblock


\bibitem[Sefelin et~al\mbox{.}(2003)]%
        {paper_protoype_2}
\bibfield{author}{\bibinfo{person}{Reinhard Sefelin}, \bibinfo{person}{Manfred Tscheligi}, {and} \bibinfo{person}{Verena Giller}.} \bibinfo{year}{2003}\natexlab{}.
\newblock \showarticletitle{Paper Prototyping - What is It Good for? A Comparison of Paper- and Computer-Based Low-Fidelity Prototyping}. In \bibinfo{booktitle}{\emph{CHI '03 Extended Abstracts on Human Factors in Computing Systems}} (Ft. Lauderdale, Florida, USA) \emph{(\bibinfo{series}{CHI EA '03})}. \bibinfo{publisher}{Association for Computing Machinery}, \bibinfo{address}{New York, NY, USA}, \bibinfo{pages}{778–779}.
\newblock
\showISBNx{1581136374}
\urldef\tempurl%
\url{https://doi.org/10.1145/765891.765986}
\showDOI{\tempurl}


\bibitem[Snyder(2003)]%
        {paper_prototype_3}
\bibfield{author}{\bibinfo{person}{Carolyn Snyder}.} \bibinfo{year}{2003}\natexlab{}.
\newblock \bibinfo{booktitle}{\emph{Paper prototyping: The fast and easy way to design and refine user interfaces}}.
\newblock \bibinfo{publisher}{Morgan Kaufmann}.
\newblock


\bibitem[Strobelt et~al\mbox{.}(2023)]%
        {prompt_ide}
\bibfield{author}{\bibinfo{person}{Hendrik Strobelt}, \bibinfo{person}{Albert Webson}, \bibinfo{person}{Victor Sanh}, \bibinfo{person}{Benjamin Hoover}, \bibinfo{person}{Johanna Beyer}, \bibinfo{person}{Hanspeter Pfister}, {and} \bibinfo{person}{Alexander~M. Rush}.} \bibinfo{year}{2023}\natexlab{}.
\newblock \showarticletitle{Interactive and Visual Prompt Engineering for Ad-hoc Task Adaptation with Large Language Models}.
\newblock \bibinfo{journal}{\emph{IEEE Transactions on Visualization and Computer Graphics}} \bibinfo{volume}{29}, \bibinfo{number}{1} (\bibinfo{year}{2023}), \bibinfo{pages}{1146--1156}.
\newblock
\urldef\tempurl%
\url{https://doi.org/10.1109/TVCG.2022.3209479}
\showDOI{\tempurl}


\bibitem[Subramonyam et~al\mbox{.}(2021)]%
        {protoai}
\bibfield{author}{\bibinfo{person}{Hariharan Subramonyam}, \bibinfo{person}{Colleen Seifert}, {and} \bibinfo{person}{Eytan Adar}.} \bibinfo{year}{2021}\natexlab{}.
\newblock \showarticletitle{ProtoAI: Model-Informed Prototyping for AI-Powered Interfaces}. In \bibinfo{booktitle}{\emph{26th International Conference on Intelligent User Interfaces}} (College Station, TX, USA) \emph{(\bibinfo{series}{IUI '21})}. \bibinfo{publisher}{Association for Computing Machinery}, \bibinfo{address}{New York, NY, USA}, \bibinfo{pages}{48–58}.
\newblock
\showISBNx{9781450380171}
\urldef\tempurl%
\url{https://doi.org/10.1145/3397481.3450640}
\showDOI{\tempurl}


\bibitem[Wang et~al\mbox{.}(2023)]%
        {popblends}
\bibfield{author}{\bibinfo{person}{Sitong Wang}, \bibinfo{person}{Savvas Petridis}, \bibinfo{person}{Taeahn Kwon}, \bibinfo{person}{Xiaojuan Ma}, {and} \bibinfo{person}{Lydia~B Chilton}.} \bibinfo{year}{2023}\natexlab{}.
\newblock \showarticletitle{PopBlends: Strategies for Conceptual Blending with Large Language Models}. In \bibinfo{booktitle}{\emph{Proceedings of the 2023 CHI Conference on Human Factors in Computing Systems}} (Hamburg, Germany) \emph{(\bibinfo{series}{CHI '23})}. \bibinfo{publisher}{Association for Computing Machinery}, \bibinfo{address}{New York, NY, USA}, Article \bibinfo{articleno}{435}, \bibinfo{numpages}{19}~pages.
\newblock
\showISBNx{9781450394215}
\urldef\tempurl%
\url{https://doi.org/10.1145/3544548.3580948}
\showDOI{\tempurl}


\bibitem[Wu et~al\mbox{.}(2023)]%
        {scattershot}
\bibfield{author}{\bibinfo{person}{Sherry Wu}, \bibinfo{person}{Hua Shen}, \bibinfo{person}{Daniel~S Weld}, \bibinfo{person}{Jeffrey Heer}, {and} \bibinfo{person}{Marco~Tulio Ribeiro}.} \bibinfo{year}{2023}\natexlab{}.
\newblock \showarticletitle{ScatterShot: Interactive In-Context Example Curation for Text Transformation}. In \bibinfo{booktitle}{\emph{Proceedings of the 28th International Conference on Intelligent User Interfaces}} (Sydney, NSW, Australia) \emph{(\bibinfo{series}{IUI '23})}. \bibinfo{publisher}{Association for Computing Machinery}, \bibinfo{address}{New York, NY, USA}, \bibinfo{pages}{353–367}.
\newblock
\showISBNx{9798400701061}
\urldef\tempurl%
\url{https://doi.org/10.1145/3581641.3584059}
\showDOI{\tempurl}


\bibitem[Wu et~al\mbox{.}(2022a)]%
        {promptchainer}
\bibfield{author}{\bibinfo{person}{Tongshuang Wu}, \bibinfo{person}{Ellen Jiang}, \bibinfo{person}{Aaron Donsbach}, \bibinfo{person}{Jeff Gray}, \bibinfo{person}{Alejandra Molina}, \bibinfo{person}{Michael Terry}, {and} \bibinfo{person}{Carrie~J Cai}.} \bibinfo{year}{2022}\natexlab{a}.
\newblock \showarticletitle{PromptChainer: Chaining Large Language Model Prompts through Visual Programming}. In \bibinfo{booktitle}{\emph{Extended Abstracts of the 2022 CHI Conference on Human Factors in Computing Systems}} (New Orleans, LA, USA) \emph{(\bibinfo{series}{CHI EA '22})}. \bibinfo{publisher}{Association for Computing Machinery}, \bibinfo{address}{New York, NY, USA}, Article \bibinfo{articleno}{359}, \bibinfo{numpages}{10}~pages.
\newblock
\showISBNx{9781450391566}
\urldef\tempurl%
\url{https://doi.org/10.1145/3491101.3519729}
\showDOI{\tempurl}


\bibitem[Wu et~al\mbox{.}(2022b)]%
        {ai-chains}
\bibfield{author}{\bibinfo{person}{Tongshuang Wu}, \bibinfo{person}{Michael Terry}, {and} \bibinfo{person}{Carrie~Jun Cai}.} \bibinfo{year}{2022}\natexlab{b}.
\newblock \showarticletitle{AI Chains: Transparent and Controllable Human-AI Interaction by Chaining Large Language Model Prompts}. In \bibinfo{booktitle}{\emph{Proceedings of the 2022 CHI Conference on Human Factors in Computing Systems}} (New Orleans, LA, USA) \emph{(\bibinfo{series}{CHI '22})}. \bibinfo{publisher}{Association for Computing Machinery}, \bibinfo{address}{New York, NY, USA}, Article \bibinfo{articleno}{385}, \bibinfo{numpages}{22}~pages.
\newblock
\showISBNx{9781450391573}
\urldef\tempurl%
\url{https://doi.org/10.1145/3491102.3517582}
\showDOI{\tempurl}


\bibitem[Yang et~al\mbox{.}(2018a)]%
        {ai_capabilities_3}
\bibfield{author}{\bibinfo{person}{Qian Yang}, \bibinfo{person}{Nikola Banovic}, {and} \bibinfo{person}{John Zimmerman}.} \bibinfo{year}{2018}\natexlab{a}.
\newblock \showarticletitle{Mapping Machine Learning Advances from HCI Research to Reveal Starting Places for Design Innovation}. In \bibinfo{booktitle}{\emph{Proceedings of the 2018 CHI Conference on Human Factors in Computing Systems}} (Montreal QC, Canada) \emph{(\bibinfo{series}{CHI '18})}. \bibinfo{publisher}{Association for Computing Machinery}, \bibinfo{address}{New York, NY, USA}, \bibinfo{pages}{1–11}.
\newblock
\showISBNx{9781450356206}
\urldef\tempurl%
\url{https://doi.org/10.1145/3173574.3173704}
\showDOI{\tempurl}


\bibitem[Yang et~al\mbox{.}(2019)]%
        {ml_communication_2}
\bibfield{author}{\bibinfo{person}{Qian Yang}, \bibinfo{person}{Justin Cranshaw}, \bibinfo{person}{Saleema Amershi}, \bibinfo{person}{Shamsi~T. Iqbal}, {and} \bibinfo{person}{Jaime Teevan}.} \bibinfo{year}{2019}\natexlab{}.
\newblock \showarticletitle{Sketching NLP: A Case Study of Exploring the Right Things To Design with Language Intelligence}. In \bibinfo{booktitle}{\emph{Proceedings of the 2019 CHI Conference on Human Factors in Computing Systems}} (Glasgow, Scotland Uk) \emph{(\bibinfo{series}{CHI '19})}. \bibinfo{publisher}{Association for Computing Machinery}, \bibinfo{address}{New York, NY, USA}, \bibinfo{pages}{1–12}.
\newblock
\showISBNx{9781450359702}
\urldef\tempurl%
\url{https://doi.org/10.1145/3290605.3300415}
\showDOI{\tempurl}


\bibitem[Yang et~al\mbox{.}(2018b)]%
        {scarce_ml_engineers_1}
\bibfield{author}{\bibinfo{person}{Qian Yang}, \bibinfo{person}{Alex Scuito}, \bibinfo{person}{John Zimmerman}, \bibinfo{person}{Jodi Forlizzi}, {and} \bibinfo{person}{Aaron Steinfeld}.} \bibinfo{year}{2018}\natexlab{b}.
\newblock \showarticletitle{Investigating How Experienced UX Designers Effectively Work with Machine Learning}. In \bibinfo{booktitle}{\emph{Proceedings of the 2018 Designing Interactive Systems Conference}} (Hong Kong, China) \emph{(\bibinfo{series}{DIS '18})}. \bibinfo{publisher}{Association for Computing Machinery}, \bibinfo{address}{New York, NY, USA}, \bibinfo{pages}{585–596}.
\newblock
\showISBNx{9781450351980}
\urldef\tempurl%
\url{https://doi.org/10.1145/3196709.3196730}
\showDOI{\tempurl}


\bibitem[Yang et~al\mbox{.}(2020)]%
        {ai_prototyping_hard_2}
\bibfield{author}{\bibinfo{person}{Qian Yang}, \bibinfo{person}{Aaron Steinfeld}, \bibinfo{person}{Carolyn Ros\'{e}}, {and} \bibinfo{person}{John Zimmerman}.} \bibinfo{year}{2020}\natexlab{}.
\newblock \showarticletitle{Re-Examining Whether, Why, and How Human-AI Interaction Is Uniquely Difficult to Design}. In \bibinfo{booktitle}{\emph{Proceedings of the 2020 CHI Conference on Human Factors in Computing Systems}} (Honolulu, HI, USA) \emph{(\bibinfo{series}{CHI '20})}. \bibinfo{publisher}{Association for Computing Machinery}, \bibinfo{address}{New York, NY, USA}, \bibinfo{pages}{1–13}.
\newblock
\showISBNx{9781450367080}
\urldef\tempurl%
\url{https://doi.org/10.1145/3313831.3376301}
\showDOI{\tempurl}


\bibitem[Yang et~al\mbox{.}(2018c)]%
        {ai_capabilities_4}
\bibfield{author}{\bibinfo{person}{Qian Yang}, \bibinfo{person}{Jina Suh}, \bibinfo{person}{Nan-Chen Chen}, {and} \bibinfo{person}{Gonzalo Ramos}.} \bibinfo{year}{2018}\natexlab{c}.
\newblock \showarticletitle{Grounding Interactive Machine Learning Tool Design in How Non-Experts Actually Build Models}. In \bibinfo{booktitle}{\emph{Proceedings of the 2018 Designing Interactive Systems Conference}} (Hong Kong, China) \emph{(\bibinfo{series}{DIS '18})}. \bibinfo{publisher}{Association for Computing Machinery}, \bibinfo{address}{New York, NY, USA}, \bibinfo{pages}{573–584}.
\newblock
\showISBNx{9781450351980}
\urldef\tempurl%
\url{https://doi.org/10.1145/3196709.3196729}
\showDOI{\tempurl}


\bibitem[Yang et~al\mbox{.}(2016)]%
        {ai_use_cases}
\bibfield{author}{\bibinfo{person}{Qian Yang}, \bibinfo{person}{John Zimmerman}, \bibinfo{person}{Aaron Steinfeld}, {and} \bibinfo{person}{Anthony Tomasic}.} \bibinfo{year}{2016}\natexlab{}.
\newblock \showarticletitle{Planning Adaptive Mobile Experiences When Wireframing}. In \bibinfo{booktitle}{\emph{Proceedings of the 2016 ACM Conference on Designing Interactive Systems}} (Brisbane, QLD, Australia) \emph{(\bibinfo{series}{DIS '16})}. \bibinfo{publisher}{Association for Computing Machinery}, \bibinfo{address}{New York, NY, USA}, \bibinfo{pages}{565–576}.
\newblock
\showISBNx{9781450340311}
\urldef\tempurl%
\url{https://doi.org/10.1145/2901790.2901858}
\showDOI{\tempurl}


\bibitem[Yuan et~al\mbox{.}(2022)]%
        {wordcraft}
\bibfield{author}{\bibinfo{person}{Ann Yuan}, \bibinfo{person}{Andy Coenen}, \bibinfo{person}{Emily Reif}, {and} \bibinfo{person}{Daphne Ippolito}.} \bibinfo{year}{2022}\natexlab{}.
\newblock \showarticletitle{Wordcraft: Story Writing With Large Language Models}. In \bibinfo{booktitle}{\emph{27th International Conference on Intelligent User Interfaces}} (Helsinki, Finland) \emph{(\bibinfo{series}{IUI '22})}. \bibinfo{publisher}{Association for Computing Machinery}, \bibinfo{address}{New York, NY, USA}, \bibinfo{pages}{841–852}.
\newblock
\showISBNx{9781450391443}
\urldef\tempurl%
\url{https://doi.org/10.1145/3490099.3511105}
\showDOI{\tempurl}


\end{thebibliography}


\end{document}